\documentclass[english,prx,aps, twocolumn, reprint, showpacs, showkeys]{revtex4-1}
\usepackage[T1]{fontenc}
\usepackage[latin9]{inputenc}
\usepackage{color}
\usepackage{verbatim}
\usepackage{amsmath}
\usepackage{amssymb}
\usepackage{graphicx}

\makeatletter
\usepackage{textgreek}
\usepackage{hyperref}

\makeatother

\usepackage{babel}
\begin{document}

\title{Hybrid quantum-classical modeling of quantum dot devices}

\author{Markus Kantner}
\email{kantner@wias-berlin.de}
\author{Markus Mittnenzweig}
\author{Thomas Koprucki}

\affiliation{Weierstrass Institute for Applied Analysis and Stochastics,\\
Mohrenstr. 39, 10117 Berlin, Germany}

\begin{abstract}
The design of electrically driven quantum dot devices for quantum
optical applications asks for modeling approaches combining classical
device physics with quantum mechanics. We connect the well-established
fields of semi-classical semiconductor transport theory and the theory
of open quantum systems to meet this requirement. By coupling the
van Roosbroeck system with a quantum master equation in Lindblad form,
we introduce a new hybrid quantum-classical modeling approach, which provides a comprehensive description of quantum dot devices on multiple scales:
It enables the calculation of quantum
optical figures of merit and the spatially resolved
simulation of the current flow in realistic semiconductor device geometries in a unified way. We construct
the interface between both theories in such a way, that the resulting hybrid
system obeys the fundamental axioms of (non-)equilibrium thermodynamics.
We show that our approach guarantees the conservation of charge, consistency
with the thermodynamic equilibrium and the second law of thermodynamics.
The feasibility of the approach is demonstrated by numerical simulations
of an electrically driven single-photon source based on a single quantum
dot in the stationary and transient operation regime. 

\end{abstract}




\maketitle

\section{Introduction}

Semiconductor quantum dots (QDs) are zero-dimensional nanostructures
which provide a discrete spectrum of electronic states due to the
confinement of charge carriers in all spatial dimensions. Because
of their tunable electro-optical properties and their easy integration
into dielectric microcavities, QDs have attracted considerable attention
in particular for applications in solid-state based optoelectronic
devices \cite{Bimberg1999,Michler2003,Bhattacharya2007,Michler2009,Bimberg2011}.
These include e.g. highly efficient semiconductor micro- and nanolasers
with a few or even a single QD as gain medium \cite{Noda2006,Gies2011,Strauf2011,Chow2013,Schneider2013},
semiconductor optical amplifiers \cite{Akiyama2007}, and quantum
light sources such as single-photon emitters and sources of entangled
photon pairs \cite{Michler2000,Santori2010,Buckley2012,Lodahl2015}.
Applications comprise optical communication and quantum information
processing \cite{Kimble2008,Santori2010,Buckley2012}, quantum cryptography
\cite{Gisin2002}, optical computing \cite{Knill2001} and bio-chemical
sensing \cite{Kairdolf2013}.

Currently, quantum optics is making the leap from the lab to commercial applications. On this way, device engineers will need simulation tools, which combine classical device physics with models from  quantum mechanics. 
The modeling and simulation of electrically driven semiconductor devices
containing QDs constitutes
a considerable challenge. On the one hand, modern optoelectronic
devices increasingly employ quantum optical effects based on coherent
light matter interaction, entanglement, photon counting statistics
and non-classical correlations, which require a quantum mechanical
description of the charge carriers and the optical field. In the last
decades, light emitting devices based on a single or a few QDs have
been successfully described by quantum master equations (QMEs) for
the density matrix \cite{Gies2011,Steinhoff2012,Chow2013},
which enable a detailed description of the dynamics of open quantum
systems. On the other hand, the simulation of electrically driven
devices requires a spatially resolved description of the current injection
from the highly doped barriers and metal contacts into the optically
active region containing the semiconductor QDs. The carrier
transport problem is well described by semi-classical transport models
such as the van Roosbroeck system \cite{VanRoosbroeck1950},
which describes the drift and diffusion of carriers within their self-consistently
generated electric field. The van Roosbroeck system has been applied
previously to QD devices, in particular to QD-based intermediate band solar cells \cite{Marti2002,Gioannini2013} and
for the optimization of the current injection in single-photon sources \cite{Kantner2016a}.

Both fields, the theory of open quantum systems and the semi-classical
semiconductor transport theory, are well developed and established
for several decades. The scope of this paper is the self-consistent
coupling of both theories in order to obtain a comprehensive description of QD-based optoelectronic
devices on multiple scales. Therefore, the interface connecting both
systems will be constructed in such a way, that the resulting hybrid
quantum-classical model guarantees the conservation of charge, consistency
with the thermodynamic equilibrium and the second law of thermodynamics.

The paper is organized as follows: In Sec.~\ref{sec:Model-equations}
the model equations are introduced and the physical properties of
the hybrid quantum-classical model are discussed. We present the structure
of the coupling terms between both systems and investigate important
features such as the conservation of charge. In Sec.~\ref{sec:Thermodynamics}
the consistency of the model equations with fundamental axioms of
$\text{(non-)}$equilibrium thermodynamics is investigated. In particular,
we construct the thermodynamic equilibrium solution by minimizing
the grand potential of the coupled system and show that the hybrid
model obeys the second law of thermodynamics. In Sec.~\ref{sec:Application}
the approach is applied to the simulation of an electrically driven single-photon
source based on a single QD. We study the stationary and transient
excitation regime by numerical simulations and show how the model
allows to compute the decisive quantum optical figures of merit along
with the spatially resolved carrier transport characteristics. 
Finally, in Sec.~\ref{sec:Outlook} we give an outlook on extensions of the approach.

\section{Model equations\label{sec:Model-equations}}

\begin{figure}
\includegraphics[width=1\columnwidth]{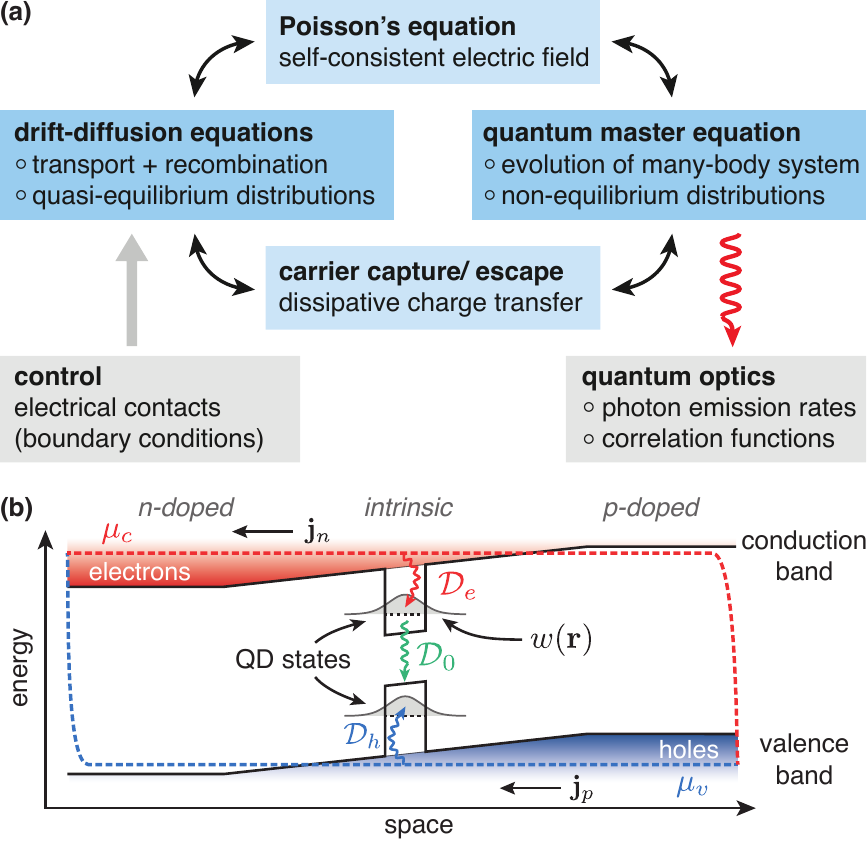}\caption{(a)~Illustration of the hybrid quantum-classical modeling approach.
A quantum system  described by a QME is self-consistently coupled to
the semi-classical transport equations for the freely roaming continuum
carriers. Both \mbox{(sub-)}systems exchange charge by capture and
escape of carriers and interact via their self-consistently generated electric
field.
(b)~Schematic band diagram of the hybrid system in a 1D cross-section of a p-i-n diode with a single QD embedded in the intrinsic zone. The dissipative interactions of the quantum system with its classical environment are described by dissipation superoperators of Lindblad type $\mathcal{D}_e$, $\mathcal{D}_h$ and $\mathcal{D}_0$ for carrier capture and escape, recombination etc. The interaction domain is determined by the spatial profile $w$.
}

\label{fig: model scheme}
\end{figure}

We consider a hybrid quantum-classical model that self-consistently
couples semi-classical transport theory to a kinetic equation for
the quantum mechanical density matrix. The latter one is a QME in
a Born-Markov and secular (rotating wave) approximation that describes
the evolution of an open quantum system which interacts with its macroscopic
environment \cite{Davies1974,Lindblad1976,Gorini1976,Breuer2002}.
In the following, the open quantum system is given by a single or
a few QDs. Our approach is based on the assumption that the charge
carriers can be separated into (free) continuum carriers and (bound)
carriers confined to QDs, which is typically met for optoelectronic
devices operating close to flat band conditions (weak electric fields)
\cite{Grupen1998,Steiger2008,Koprucki2011}. The model equations read
\begin{align}
-\nabla\cdot\varepsilon\nabla\psi & =q\left(p-n+C+Q\left(\rho\right)\right),\label{eq: Poisson equation}\\
\partial_{t}n-\frac{1}{q}\nabla\cdot\mathbf{j}_{n} & =-R-S_{n}\left(\rho;n,p,\psi\right),\label{eq: electron transport}\\
\partial_{t}p+\frac{1}{q}\nabla\cdot\mathbf{j}_{p} & =-R-S_{p}\left(\rho;n,p,\psi\right),\label{eq: hole transport}\\
\frac{\mathrm{d}}{\mathrm{d}t}\rho = \mathcal{L}\left(\rho;n,p,\psi\right)  &=-\frac{i}{\hbar}\left[H,\rho\right]+\mathcal{D}\left(\rho;n,p,\psi\right)\label{eq: quantum master equation}
\end{align}
on the domain $\Omega\subset\mathbb{R}^{3}$. The system (\ref{eq: Poisson equation})--(\ref{eq: quantum master equation})
is subject to initial conditions and boundary conditions modeling
electrical contacts and other interfaces \cite{Selberherr1984}. See Appendix~\ref{sec:Boundary-conditions}
for the boundary conditions considered throughout this paper. A schematic
illustration of the modeling approach is shown in Fig.~\ref{fig: model scheme}.

The model (\ref{eq: Poisson equation})--(\ref{eq: quantum master equation}) differs from the typical quantum optical setting by explicitly considering the spatially resolved semi-classical carrier transport equations (\ref{eq: Poisson equation})--(\ref{eq: hole transport}) as a part of the system under investigation.
As a consequence, here the notion ``reservoir'' is employed differently from the standard quantum optics literature.
In the following, the term \emph{reservoir} refers to the electrical contacts connected to the semiconductor device and the surrounding heat bath, which must be distinguished from the \emph{classical} or \emph{macroscopic environment} of the quantum system, see Fig.~\ref{fig: system-reservoir}.
The continuum carriers, which represent the electronic part of the classical environment of the quantum system,  evolve according to the van Roosbroeck system.

\subsection{Van Roosbroeck system}

Eqns.~(\ref{eq: Poisson equation})--(\ref{eq: hole transport})
represent the standard van Roosbroeck system, extended by additional
terms that constitute the coupling to the quantum system. Poisson's
Eq.~(\ref{eq: Poisson equation}) describes the electrostatic potential
$\psi$ generated by the free electron and hole densities $n$ and
$p$, the (stationary) built-in doping profile $C$ and the expectation
value of the charge density $Q\left(\rho\right)$ of the carriers
confined to the QDs. The dielectric permittivity of the semiconductor
material is given by $\varepsilon=\varepsilon_{0}\varepsilon_{r}$
and $q$ denotes the elementary charge. The continuity equations (\ref{eq: electron transport})--(\ref{eq: hole transport})
describe the flux of free electrons and holes in the presence of recombination
and transitions between free and bound states. The (net-)recombination
rate $R$ includes several recombination channels such as Shockley-Read-Hall
recombination, spontaneous emission and Auger recombination. Moreover,
carriers can be scattered from the continuum to the QDs which is described
by the ${\text{(net-)}}$capture rates $S_{n}$ and $S_{p}$. The
van Roosbroeck system must be augmented with additional state equations
for the free carrier densities
\begin{subequations}\label{eq: carrier densities}
\begin{align}
n & =N_{c}F_{1/2}\left(\beta\left(\mu_{c}-E_{c}+q\psi\right)\right),\label{eq: electron density}\\
p & =N_{v}F_{1/2}\left(\beta\left(E_{v}-q\psi-\mu_{v}\right)\right)\label{eq: hole density}
\end{align}
\end{subequations}
and the electrical current densities
\begin{subequations}\label{eq: current densities}
\begin{align}
\mathbf{j}_{n} & =\frac{1}{q}\sigma_{n}\nabla\mu_{c},\label{eq: electron current density}\\
\mathbf{j}_{p} & =\frac{1}{q}\sigma_{p}\nabla\mu_{v}.\label{eq: hole current density}
\end{align}
\end{subequations}
Here, $N_{c}$ and $N_{v}$ denote the effective
density of states of the conduction and valence band and $E_{c}$
and $E_{v}$ are the respective band edge energies. The inverse temperature
$\beta=\left(k_{B}T\right)^{-1}$ is considered as a fixed parameter
and
\[
F_{\nu}\left(\eta\right)=\frac{1}{\Gamma\left(\nu+1\right)}\int_{0}^{\infty}\mathrm{d}\xi\,\frac{\xi^{\nu}}{e^{\xi-\eta}+1}
\]
is the Fermi--Dirac integral of order $\nu$. The state
equations (\ref{eq: carrier densities}) describe thermalized carrier
ensembles in a quasi-equilibrium distribution, where the quasi-Fermi
energies of the conduction band $\mu_{c}$ and the valence band $\mu_{v}$
parametrize the deviation from the thermodynamic equilibrium.
 In accordance with linear irreversible thermodynamics, the current
densities are driven by the gradients of the quasi-Fermi energies
\cite{DeGroot1984}. The electrical conductivities $\sigma_{n}=qM_{n}n$,
$\sigma_{p}=qM_{p}p$ are products of the carrier densities and
the carrier mobilities $M_{n/p}$.

\subsection{Quantum master equation}

The state of the quantum system is described by the density matrix
$\rho$, which is subject to the QME~(\ref{eq: quantum master equation}).
 Here, the quantum system represents a many-body problem describing
the charge carriers confined to QDs and possibly further quasi-particles,
e.g. cavity photons, phonons or exciton-polaritons (dressed states). 

The Hamiltonian in Eq.~(\ref{eq: quantum master equation}) takes
the form 
\[
H=H_{0}+H_{I},
\]
where $H_{0}$ describes the single-particle energies of the confined electrons
and holes (and possibly additional particle species).
The interaction Hamiltonian $H_{I}$ is assumed to commute with the
charge number operator of the quantum system
\begin{equation}
N=n_{e}-n_{h}\label{eq: net charge operator}
\end{equation}
($n_{e}$ and $n_{h}$ are the number operators of the bound electrons
and holes) such that the Hamiltonian part of the evolution conserves
the net charge
\begin{equation}
\left[H,N\right]=0.\label{eq: commutator N and H}
\end{equation}
This imposes only a weak restriction on $H_{I}$ and allows e.g. for
Coulomb interaction between the confined carriers (configuration interaction)
as well as coherent light-matter interaction.

We assume the quantum system to be embedded in a semiconductor device, which represents a macroscopic environment with an infinitely large number of degrees of freedom. The interactions of the quantum system with its environment, e.g. the exchange of energy and charge via recombination and capture or escape of carriers, represent dissipative processes that are described by the dissipation superoperator $\mathcal{D}$. Within the limit of weak system-reservoir coupling one obtains by using the Born-Markov and secular (rotating wave) approximation a dissipation superoperator in Lindblad form \cite{Breuer2002, Schaller2014}
\begin{align}
\begin{aligned}
\mathcal{D}\left(\rho;\chi\right) &= \sum_{\alpha\in I_{\alpha}} \mathcal{D}_{\alpha }\left(\rho;\chi\right) \\
& =\sum_{\alpha\in I_{\alpha}}\big(\gamma_{\alpha}(\chi)L_{A_{\alpha}}(\rho)+\hat{\gamma}_{\alpha}(\chi) L_{A^{\dagger}_{\alpha}}(\rho)\big)
\end{aligned}
\label{eq: Dissipator (quantum detailed balance)}
\end{align}
with the \emph{Lindblad superoperator}
\begin{align*}
L_{A}\left(\rho\right)=A\rho A^{\dagger}-\frac{1}{2}\left\{ A^{\dagger}A,\rho\right\}.
\end{align*}
The admitted irreversible interactions between the quantum system and its environment are indexed by $\alpha\in I_{\alpha}$.
The environment considered in this paper is a tensor product of multiple thermal states. This comprises a bosonic heat bath (lattice phonons, thermal radiation) and the thermalized carrier ensembles, which are subject to the van Roosbroeck system (\ref{eq: Poisson equation})--(\ref{eq: hole transport}).
In the hybrid model, the forward and backward transition rates $\gamma_{\alpha}$ and $\hat{\gamma}_{\alpha}$ depend on the state of the macroscopic environment, which is indicated here by the state vector $\chi$.
Under the assumptions and approximations outlined above, the dissipation superoperator can be additively decomposed into various channels as given in Eq.~(\ref{eq: Dissipator (quantum detailed balance)}) \cite{Schaller2014}. 
A QME in Lindblad form ensures the preservation of trace, hermiticity
and (complete) positivity of the density matrix \cite{Lindblad1976,Gorini1976}.
The symbol ${\left\{ A,B\right\} =AB+BA}$ denotes the anti-commutator.
The operators $A_{\alpha}$ represent the \emph{quantum jump operators},
which are projectors between different eigenstates of $H$.
Following the standard construction of a Lindblad-QME for a weak system-reservoir
interaction \cite{Breuer2002} (extended to the case of
variable charge number here), we require the jump operators to satisfy
\begin{subequations}\label{eq: jump operator requirements}
\begin{align}
\left[H,A_{\alpha}\right] & =-\hbar\omega_{\alpha}A_{\alpha},\label{eq: jump operator requirement H}\\
\left[N,A_{\alpha}\right] & =-\ell_{\alpha}A_{\alpha},\label{eq: jump operator requirement N}
\end{align}
\end{subequations}
where $\hbar\omega_{\alpha}$ denotes the transition
energy and ${\ell_{\alpha}\in\mathbb{Z}}$ quantifies the charge transfer
of the interaction described by $A_{\alpha}$.
In order to classify the dissipation superoperators with respect to their effect on the charge of the quantum system, we collect
the dissipators belonging to equal values of $\ell_{\alpha}$
and introduce the notation
\begin{equation}
\mathcal{D}\left(\rho;\chi\right)=\mathcal{D}_{e}\left(\rho;\chi\right)+\mathcal{D}_{h}\left(\rho;\chi\right)+\mathcal{D}_{0}\left(\rho;\chi\right), \label{eq: dissipator decomposition}
\end{equation}
where we have split the index set $I_{\alpha}$ into three disjoint subsets $I_{\alpha} = I_{e} \cup I_{h} \cup  I_{0}$.
With $\ell_{\alpha\in I_e}=-1$ and $\ell_{\alpha\in I_h}=+1$, the dissipators $\mathcal{D}_{e}$ and $\mathcal{D}_{h}$
can change the charge of the quantum system
(by capture and escape of electrons and holes), whereas the processes
described by $\mathcal{D}_{0}$ with $\ell_{\alpha\in I_0}=0$ leave the charge
invariant (e.g. spontaneous emission, photon absorption, intraband
carrier relaxation, outcoupling of cavity photons). Simultaneous capture of multiple carriers with $\vert\ell_{\alpha}\vert \geq 2$ is neglected here.
From Eq.~(\ref{eq: jump operator requirement N}) and $\ell_{\alpha\in I_0} = 0$ one easily obtains
\begin{equation}
\mathrm{tr}\left(N\mathcal{D}_{0}\left(\rho;\chi\right)\right)=0.\label{eq: charge conservation of D0}
\end{equation}
Throughout this paper, we restrict ourselves to dissipation superoperators which
satisfy the quantum detailed balance condition with respect to the
thermodynamic equilibrium \cite{Alicki1976,Kossakowski1977}. This requires a certain relationship between the forward and backward transition rates $\gamma_{\alpha}$ and $\hat{\gamma}_{\alpha}$, which will be discussed in Sec.~\ref{subsec:Microscopic-transition-rates}.
In the case of degenerate energy spectra, the traditional secular approximation must be modified to properly account for degenerate eigenstate coherences. As shown in \cite{Cuetara2016}, this can be done in a thermodynamically consistent way.
Finally, we remark that the Lamb-Shift is neglected throughout this paper.

\subsection{Macroscopic coupling terms and charge conservation\label{subsec:Coupling-terms-and}}

By taking the time derivative of Poisson's Eq.~(\ref{eq: Poisson equation})
and using Eq.~(\ref{eq: electron transport})--(\ref{eq: hole transport}),
we obtain the continuity equation
\begin{align*}
\nabla\cdot\mathbf{j}_{\text{tot}} & =q\left(\partial_{t}Q-S_{p}+S_{n}\right)
\end{align*}
for the total current density $\mathbf{j}_{\text{tot}}=\mathbf{j}_{n}+\mathbf{j}_{p}+\partial_{t}\mathbf{D}$.
Besides the flux of charge carriers, it also includes the displacement
current density $\partial_{t}\mathbf{D}=-\varepsilon\partial_{t}\nabla\psi$.
For the sake of simplicity, we consider a quantum system comprising
only a single QD. The generalization of the approach outlined below
to the case of multiple QDs is straightforward. We approximate the
electric charge density of the QD by the expectation value of the
(net-)charge operator
\begin{equation}
Q\left(\rho\right)=-w\left(\mathbf{r}\right)\mathrm{tr}\left(N\rho\right),\label{eq: charge density of the quantum system}
\end{equation}
where $w$ models the spatial profile of the captured carriers, which
is assumed to be identical for all carriers. The function $w$ is
normalized such that $\int_{\Omega}\mathrm{d}^{3}r\,w\left(\mathbf{r}\right)=1$.
The spatial profile $w$ replaces the absolute squares of the many-body
wave functions of the bound carriers. The actual spatial distributions
of the confined carriers differ only on a small length scale, which
can be safely neglected in the simulation of macroscopic charge transport.
In the form of Eq.~(\ref{eq: charge density of the quantum system}),
the model accounts for long range electrostatic correlations induced
by the confined carriers. 

Using Eqns.~(\ref{eq: quantum master equation}), (\ref{eq: commutator N and H}),
(\ref{eq: dissipator decomposition}) and (\ref{eq: charge conservation of D0}),
the time derivative of Eq.~(\ref{eq: charge density of the quantum system})
is obtained as
\begin{align*}
\partial_{t}Q &= -w\left(\mathbf{r}\right) \mathrm{tr}\left(N\mathcal{D}_{e}\left(\rho;n,p,\psi\right)\right)\\
&\phantom{=}\; -w\left(\mathbf{r}\right) \mathrm{tr}\left(N\mathcal{D}_{h}\left(\rho;n,p,\psi\right)\right).
\end{align*}
In order to ensure local charge conservation  $\nabla\cdot\mathbf{j}_{\text{tot}}=0$,
the (net-)capture rates appearing in the carrier transport equations (\ref{eq: electron transport})
and (\ref{eq: hole transport}) are identified as\begin{subequations}\label{eq: loss terms}
\begin{align}
S_{n} & =+w\left(\mathbf{r}\right)\mathrm{tr}\left(N\mathcal{D}_{e}\left(\rho;n,p,\psi\right)\right),\label{eq: electron loss term}\\
S_{p} & =-w\left(\mathbf{r}\right)\mathrm{tr}\left(N\mathcal{D}_{h}\left(\rho;n,p,\psi\right)\right).\label{eq: hole loss term}
\end{align}
\end{subequations}The (net-)capture rates $S_{n/p}$ contain all
microscopic capture processes connected with transitions between the
various multi-particle configurations of the QD.

For different choices of $Q\left(\rho\right)$, e.g. different localization
profiles of captured electrons and holes $Q\left(\rho\right)=w_{h}\left(\mathbf{r}\right)\mathrm{tr}\left(n_{h}\rho\right)-w_{e}\left(\mathbf{r}\right)\mathrm{tr}\left(n_{e}\rho\right)$
(with $w_{e/h}$ normalized), the property of local charge
conservation is lost in general. However, the violation of local charge
conservation is restricted to a small region $\nabla\cdot\mathbf{j}_{\text{tot}}\propto\left(w_{e}\left(\mathbf{r}\right)-w_{h}\left(\mathbf{r}\right)\right)$
and is preserved globally, i.e. it holds $\int_{\Omega}\mathrm{d}^{3}r\,\nabla\cdot\mathbf{j}_{\text{tot}}=0$.

The thermodynamic consistency discussed in the subsequent sections does not crucially rely on the property of \emph{local} charge conservation as enforced by Eq.~(\ref{eq: loss terms}). With some minor modifications, the approach can be generalized to cases where only the weaker condition of global charge conservation is fulfilled.
This allows e.g. for capture rates with a more complicated spatial dependency than the one stated in Eq.~(\ref{eq: loss terms}).
Since the discussion of thermodynamic consistency is least technical in the case of local charge conservation, we assume Eq.~(\ref{eq: loss terms}) in the following. Other cases can be treated analogously.

\begin{figure}
\includegraphics[width=1.0\columnwidth]{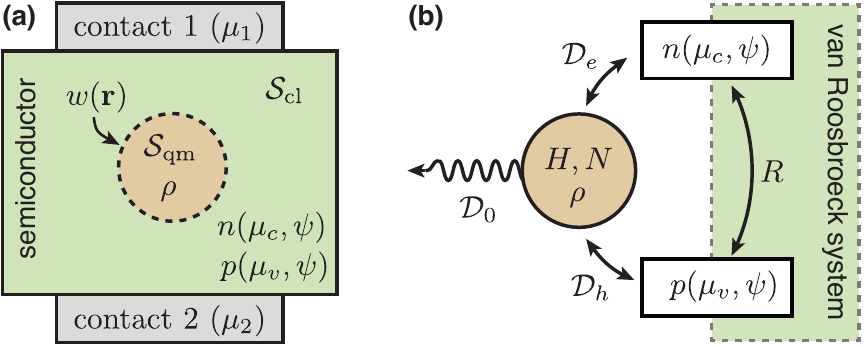}
\caption{
(a)~Spatial arrangement of the system $\mathcal{S}$ and the electrical contacts (reservoirs). 
The system $\mathcal{S}$
consists of a classical subsystem $\mathcal{S}_{\text{cl}}$ and
a quantum mechanical subsystem $\mathcal{S}_{\text{qm}}$ in the interior
of the classical domain. The classical system is in contact with several electric contacts, which act as charge reservoirs and are characterized by their chemical potentials $\mu_i$ and a common inverse temperature $\beta$. The reservoirs enter the equations via Dirichlet boundary conditions.
(b)~Illustration of the coupling scheme. The quantum system is coupled to the van Roosbroeck system via charge transfer mediated by dissipation superoperators $\mathcal{D}_e$ and $\mathcal{D}_h$. Further decay processes, which keep the charge of the quantum system invariant, are described by $\mathcal{D}_0$. Besides the charge transfer, the model system (\ref{eq: Poisson equation})--(\ref{eq: quantum master equation}) accounts for electrostatic interaction between charges in $\mathcal{S}_{\text{cl}}$ and $\mathcal{S}_{\text{qm}}$.
}

\label{fig: system-reservoir}
\end{figure}

\section{Thermodynamics\label{sec:Thermodynamics}}

In the recent years, the on-going miniaturization of (quantum) electronic
devices has enabled the investigation of thermodynamical laws on the
nanoscale. This has lead to the emergence of the novel field of \emph{quantum
thermodynamics} \cite{Gemmer2004,Kosloff2013,Esposito2015,Goold2016}.
Experiments and theory indicate that the fundamental thermodynamical
laws also hold in the quantum regime \cite{Pekola2015,Strasberg2017}
and therefore we view thermodynamic consistency as a crucial feature
for any hybrid quantum-classical model.

In this section we discuss the thermodynamic properties of the hybrid
model system (\ref{eq: Poisson equation})--(\ref{eq: quantum master equation}).
At first, this concerns a consideration of the energy, charge and
entropy balance between the system and its reservoirs.  Second, the
thermodynamic equilibrium solution of the hybrid system will be constructed
by minimizing its grand potential. Moreover, we formulate a relation
between the microscopic transition rates satisfying the quantum detailed
balance condition. Finally, the hybrid quantum-classical model (\ref{eq: Poisson equation})--(\ref{eq: quantum master equation})
is shown to have a non-negative entropy production rate, which we
interpret as consistency with the second law of thermodynamics.

\subsection{Energy, charge and entropy balance}

We consider an open system $\mathcal{S}$, which itself consists of
a classical subsystem $\mathcal{S}_{\text{cl}}$ and quantum-mechanical subsystem $\mathcal{S}_{\text{qm}}$. The system $\mathcal{S}$ is in contact with several reservoirs $\mathcal{R}_{i}$ as illustrated in Fig.~\ref{fig: system-reservoir}(a). The system $\mathcal{S}$
can exchange energy and charge carriers with the reservoirs. The combined
system is assumed to be isolated. The
reservoir $\mathcal{R}_{0}$ is a heat bath with fixed background temperature $T$, which comprises the crystal lattice as well
as the surrounding radiation field. The reservoirs $\mathcal{R}_{i\geq1}$
model the electrical contacts at the boundary of the device. They are
characterized by a common temperature and their chemical potentials $\mu_{i}$ (or applied voltages), which enter the system (\ref{eq: Poisson equation})--(\ref{eq: quantum master equation}) via boundary conditions (cf. Appendix \ref{sec:Boundary-conditions}).

The total change of entropy is given by
\[
\Delta S_{\text{tot}}=\Delta S_{\mathcal{S}}+\Delta S_{\mathcal{R}}\geq0,
\]
and the conservation of the total internal energy and charge is expressed
as
\begin{align*}
\Delta U & =\Delta U_{\mathcal{S}}+\Delta U_{\mathcal{R}}=0,\\
\Delta N & =\Delta N_{\mathcal{S}}+\Delta N_{\mathcal{R}}=0.
\end{align*}
The reservoir $\mathcal{R}_{0}$ can exchange only energy with $\mathcal{S}$,
hence its change of entropy is given by $\Delta S_{\mathcal{R}_{0}}=\frac{1}{T}\Delta U_{\mathcal{R}_{0}}$.
For the contacts $\mathcal{R}_{i\geq1}$, also charge transfer is
possible such that $\Delta S_{\mathcal{R}_{i\geq 1}}=\frac{1}{T}\Delta U_{\mathcal{R}_{i}}-\frac{\mu_{i}}{T}\Delta N_{\mathcal{R}_{i}}$.
Using the conservation laws state above and $\Delta U_{\mathcal{R}} = \sum_{i\geq 0}\Delta U_{\mathcal{R}_i}$, we obtain
\[
\Delta S_{\text{tot}}=\Delta S_{\mathcal{S}}-\frac{1}{T}\Delta U_{\mathcal{S}}-\sum_{i\geq1}\frac{\mu_{i}}{T}\Delta N_{\mathcal{R}_{i}},
\]
where $\Delta N_{\mathcal{R}_{i}}$ is just the (negative) charge flow
across the boundary $\Gamma_{i}$. Using 
\[
\lim_{\Delta t\to0}\frac{\mathrm{\Delta}N_{\mathcal{R}_{i}}}{\Delta t}=\frac{\mathrm{d}N_{\mathcal{R}_{i}}}{\mathrm{d}t}=-\frac{1}{q}\int_{\Gamma_{i}}\mathrm{d}\mathbf{A}\cdot\left(\mathbf{j}_{n}+\mathbf{j}_{p}\right),
\]
we obtain the entropy production rate
\begin{equation}
\frac{\mathrm{d}S_{\text{tot}}}{\mathrm{d}t}=-\frac{1}{T}\frac{\mathrm{d}F_{\mathcal{S}}}{\mathrm{d}t}+\sum_{i\geq1}\frac{\mu_{i}}{qT}\int_{\Gamma_{i}}\mathrm{d}\mathbf{A}\cdot\left(\mathbf{j}_{n}+\mathbf{j}_{p}\right),\label{eq: entropy production (general)}
\end{equation}
where $F_{\mathcal{S}}=U_{\mathcal{S}}-TS_{\mathcal{S}}$ denotes
the free energy of the system $\mathcal{S}$. In Sec.~\ref{subsec:Entropy-production-and}
it will be shown, that the entropy production rate is indeed always
positive for the hybrid model (\ref{eq: Poisson equation})--(\ref{eq: quantum master equation}).
Under chemical equilibrium boundary conditions (all reservoirs $\mathcal{R}_{i\geq1}$
have the chemical potential $\mu_{i}=\mu_{\text{eq}}$), the
above expression simplifies further. Exploiting the conservation of
total charge, one obtains
\begin{equation}
\left.\frac{\mathrm{d}S_{\text{tot}}}{\mathrm{d}t}\right\vert _{\text{eq}}=-\frac{1}{T}\frac{\mathrm{d}\Omega_{\mathcal{S}}}{\mathrm{d}t}\label{eq: entropy production (equilibrium conditions)}
\end{equation}
with the grand potential $\Omega_{\mathcal{S}}=U_{\mathcal{S}}-TS_{\mathcal{S}}-\mu_{\text{eq}}N_{\mathcal{S}}$.
Thus, $\Omega_{\mathcal{S}}$ is a Lyapunov function
for the irreversible relaxation of $\mathcal{S}$ into the thermodynamic
equilibrium. 

\subsection{Thermodynamic equilibrium\label{subsec:Thermodynamic-equilibrium} }

According to Eq.~(\ref{eq: entropy production (equilibrium conditions)}),
the thermodynamic equilibrium solution of (\ref{eq: Poisson equation})--(\ref{eq: quantum master equation})
can be constructed by minimizing the grand potential $\Omega_{\mathcal{S}}$.
Since we assume only a weak coupling between the quantum system and
its macroscopic environment, the total entropy, total internal energy
and total charge number are given by sums of the classical and the
quantum mechanical contribution
\begin{subequations}\label{eq: total thermodynamic potentials}
\begin{align}
S\left(n,p,\rho\right) & =S_{\text{cl}}\left(n,p\right)+S_{\text{qm}}\left(\rho\right),\label{eq: total entropy}\\
U\left(n,p,\rho\right) & =U_{\text{cl}}\left(n,p\right)+U_{\text{qm}}\left(\rho\right)\nonumber \\
 & \phantom{=}+U_{\psi}\left(p-n+Q\left(\rho\right)\right),\label{eq: total internal energy}\\
N\left(n,p,\rho\right) & =N_{\text{cl}}\left(n,p\right)+N_{\text{qm}}\left(\rho\right).\label{eq: total charge number}
\end{align}
\end{subequations}
Here also the energy contribution $U_{\psi}$
of the electrostatic field is taken into account. The extensive thermodynamic
quantities of the macroscopic system are expressed via volume densities
\begin{align*}
S_{\text{cl}}\left(n,p\right) & =\int_{\Omega}\mathrm{d}^{3}r\,s_{\text{cl}}\left(n,p\right),\\
U_{\text{cl}}\left(n,p\right) & =\int_{\Omega}\mathrm{d}^{3}r\,u_{\text{cl}}\left(n,p\right),\\
N_{\text{cl}}\left(n,p\right) & =\int_{\Omega}\mathrm{d}^{3}r\,\left(n-p\right)
\end{align*}
with the entropy density $s_{\text{cl}}$ and the internal energy
density $u_{\text{cl}}$. We consider the continuum carriers to be
in a \emph{local thermodynamic equilibrium} \cite{DeGroot1984}. Hence,
the internal energy density and the entropy density can be expressed
as functions of the local carrier density
\begin{subequations}\label{eq: classical entropy and energy density}
\begin{align}
s_{\text{cl}} & =-k_{B}\left(nF_{1/2}^{-1}\left(\frac{n}{N_{c}}\right)-\frac{5}{2}N_{c}F_{3/2}\left(F_{1/2}^{-1}\left(\frac{n}{N_{c}}\right)\right)\right)\nonumber \\
 & \phantom{=}-k_{B}\left(pF_{1/2}^{-1}\left(\frac{p}{N_{v}}\right)-\frac{5}{2}N_{v}F_{3/2}\left(F_{1/2}^{-1}\left(\frac{p}{N_{v}}\right)\right)\right),\label{eq: classical entropy density}\\
u_{\text{cl}} & =\frac{3}{2}k_{B}TN_{c}F_{3/2}\left(F_{1/2}^{-1}\left(\frac{n}{N_{c}}\right)\right)+E_{c}n\nonumber \\
 & \phantom{=}+\frac{3}{2}k_{B}TN_{v}F_{3/2}\left(F_{1/2}^{-1}\left(\frac{p}{N_{v}}\right)\right)-E_{v}p.\label{eq: classical energy density}
\end{align}
\end{subequations}
The above relations are obtained for the quasi-free
electron and hole gas with parabolic energy dispersion and Fermi--Dirac
statistics in three dimensions \cite{Albinus2002}. The contributions
of the quantum system are given by the von Neumann entropy and the
expectation values of the Hamiltonian $H$ and the charge number operator
$N$\begin{subequations}\label{eq: quantum mechanical entropy and energy density}
\begin{align}
S_{\text{qm}} & =-k_{B}\mathrm{tr}\left(\rho\log{\rho}\right),\label{eq: quantum entropy}\\
U_{\text{qm}} & =\mathrm{tr}\left(H\rho\right),\label{eq: quantum internal energy}\\
N_{\text{qm}} & =\mathrm{tr}\left(N\rho\right).\label{eq: quantum charge}
\end{align}
\end{subequations}
The carriers interact via their self-consistently
generated electrostatic field, which yields the contribution $U_{\psi}$
to the internal energy. It is convenient to decompose the total electrostatic
potential into $\psi=\psi_{\text{int}}+\psi_{\text{ext}}$, where
the internal field $\psi_{\text{int}}=\psi_{\text{int}}\left(\rho_{\text{int}}\right)$
is generated by the total internal carrier density 
\[
\rho_{\text{int}}=p-n+Q\left(\rho\right),
\]
whereas the external field $\psi_{\text{ext}}$ arises from the built-in
doping profile and voltages applied at the electric contacts. Then,
the field energy can be written as \cite{Albinus1996}
\begin{align}
U_{\psi}\left(\rho_{\text{int}}\right) = \frac{1}{2}\int_{\Omega}\mathrm{d}^{3}r\,\varepsilon\left|\nabla\psi_{\text{int}}\left(\rho_{\text{int}}\right)\right|^{2}+q\int_{\Omega}\mathrm{d}^{3}r\,\rho_{\text{int}}\psi_{\text{ext}}.
\label{eq: internal energy electric field}
\end{align}

Assuming the charge density of the quantum system as stated in Eq.~(\ref{eq: charge density of the quantum system}),
and finally minimizing the grand potential $\Omega_{\mathcal{S}}$
under the constraint $\mathrm{tr}\left(\rho\right)=1$, we obtain
the equilibrium free carrier densities as
\begin{align*}
n_{\text{eq}} & =N_{c}F_{1/2}\left(\beta\left(\mu_{\text{eq}}-E_{c}+q\psi_{\text{eq}}\right)\right),\\
p_{\text{eq}} & =N_{v}F_{1/2}\left(\beta\left(E_{v}-q\psi_{\text{eq}}-\mu_{\text{eq}}\right)\right)
\end{align*}
and the equilibrium density matrix
\begin{equation}
\rho_{\text{eq}}=\frac{1}{Z}e^{-\beta\left(H-\left(\mu_{\text{eq}}+q\left\langle \psi_{\text{eq}}\right\rangle _{w}\right)N\right)}.\label{eq: thermal equilibrium density matrix}
\end{equation}
Here, ${Z=\mathrm{tr}\left(\exp{\left(-\beta\left(H-\left(\mu_{\text{eq}}+q\left\langle \psi_{\text{eq}}\right\rangle _{w}\right)N\right)\right)}\right)}$
represents the grand canonical partition function, 
\begin{equation}
\left\langle \psi\right\rangle _{w}=\int_{\Omega}\mathrm{d}^{3}r\,w\left(\mathbf{r}\right)\psi\left(\mathbf{r}\right)\label{eq: spatial average}
\end{equation}
is the averaged electrostatic potential in the vicinity of the QD
and the \emph{built-in potential} $\psi_{\text{eq}}$ solves Eq.~(\ref{eq: Poisson equation})
with the right hand side $q\left(p_{\text{\text{eq}}}-n_{\text{\text{eq}}}+C+Q\left(\rho_{\text{\text{eq}}}\right)\right)$
at equilibrium boundary conditions. The equilibrium density matrix
is a grand canonical ensemble, which contains a contribution from
the electrostatic potential due to the electrostatic interaction with
the macroscopic environment. The latter
appears in Eq.~(\ref{eq: thermal equilibrium density matrix}) as a spatial average using the localization profile $w$ of
the confined carriers as a weighting function, see Eq.~(\ref{eq: spatial average}).
This is a remarkable result, which indicates that the quantum system
interacts only with its spatially averaged macroscopic environment.
We emphasize that this is a direct consequence of the ansatz Eq.~(\ref{eq: charge density of the quantum system})
and the variation of Eq.~(\ref{eq: internal energy electric field})
with respect to $n$, $p$ and $\rho$. See Appendix~\ref{sec:Electrostatic-field-energy}
for details.

In the following, the concept of a non-local interaction of the quantum
system with its spatially averaged macroscopic environment will be
extended to non-equilibrium situations.

\subsection{Microscopic transition rates and the quantum detailed balance condition\label{subsec:Microscopic-transition-rates}}

We assume the microscopic transition rates in the dissipator (\ref{eq: Dissipator (quantum detailed balance)})
to be functions of the spatially averaged macroscopic potentials
\begin{align*}
\gamma_{\alpha} & =\gamma_{\alpha}\left(\left\langle \mu_{c}\right\rangle _{w},\left\langle \mu_{v}\right\rangle _{w},\left\langle \psi\right\rangle _{w}\right),\\
\hat{\gamma}_{\alpha} & =\hat{\gamma}_{\alpha}\left(\left\langle \mu_{c}\right\rangle _{w},\left\langle \mu_{v}\right\rangle _{w},\left\langle \psi\right\rangle _{w}\right),
\end{align*}
where $\left\langle \cdot\right\rangle _{w}$ denotes the spatial
average according to Eq.~(\ref{eq: spatial average}). The quantum
detailed balance condition requires the dissipator to vanish in equilibrium.
Hence, the condition
\begin{align*}
0&\stackrel{!}{=}\mathcal{D}_{\alpha}\left(\rho_{\text{eq}};n_{\text{eq}},p_{\text{eq}},\psi_{\text{eq}}\right) =\gamma_{\alpha}^{\text{eq}}L_{A_{\alpha}}(\rho_{\text{eq}}) + \hat{\gamma}_{\alpha}^{\text{eq}}L_{A^{\dagger}_{\alpha}}(\rho_{\text{eq}})
\end{align*}
can be used to derive a relation between the equilibrium transition
rates $\gamma_{\alpha}^{\text{eq}}=\gamma_{\alpha}\left(\mu_{\text{eq}},\mu_{\text{eq}},\left\langle \psi_{\text{eq}}\right\rangle _{w}\right)$
and $\hat{\gamma}_{\alpha}^{\text{eq}}$. From Eq.~(\ref{eq: jump operator requirements}),
one obtains for any $\lambda\in\mathbb{R}$
\begin{align*}
e^{\lambda H}A_{\alpha}e^{-\lambda H} & =e^{-\lambda\hbar\omega_{\alpha}}A_{\alpha},\\
e^{\lambda N}A_{\alpha}e^{-\lambda N} & =e^{-\lambda\ell_{\alpha}}A_{\alpha},
\end{align*}
which implies
\begin{align*}
A_{\alpha}\rho_{\text{eq}} & =e^{-\beta\left(\hbar\omega_{\alpha}-\left(\mu_{\text{eq}}+q\left\langle \psi_{\text{eq}}\right\rangle _{w}\right)\ell_{\alpha}\right)}\rho_{\text{eq}}A_{\alpha},\\
A_{\alpha}^{\dagger}\rho_{\text{eq}} & =e^{+\beta\left(\hbar\omega_{\alpha}-\left(\mu_{\text{eq}}+q\left\langle \psi_{\text{eq}}\right\rangle _{w}\right)\ell_{\alpha}\right)}\rho_{\text{eq}}A_{\alpha}^{\dagger}.
\end{align*}
Subsequently, one obtains
\begin{align*}
0 &\stackrel{!}{=}\left(\gamma_{\alpha}^{\text{eq}}-\hat{\gamma}_{\alpha}^{\text{eq}}e^{+\beta\left(\hbar\omega_{\alpha}-\left(\mu_{\text{eq}}+q\left\langle \psi_{\text{eq}}\right\rangle _{w}\right)\ell_{\alpha}\right)}\right)\times\\
 & \phantom{=}\times\Big(A_{\alpha}\rho_{\text{eq}}A_{\alpha}^{\dagger}-e^{-\beta\left(\hbar\omega_{\alpha}-\left(\mu_{\text{eq}}+q\left\langle \psi_{\text{eq}}\right\rangle _{w}\right)\ell_{\alpha}\right)}A_{\alpha}^{\dagger}\rho_{\text{eq}}A_{\alpha}\Big),
\end{align*}
which yields the desired relation between $\gamma_{\alpha}^{\text{eq}}$
and $\hat{\gamma}_{\alpha}^{\text{eq}}$:
\[
\hat{\gamma}_{\alpha}^{\text{eq}}=\gamma_{\alpha}^{\text{eq}}e^{-\beta\left(\hbar\omega_{\alpha}-\left(\mu_{\text{eq}}+q\left\langle \psi_{\text{eq}}\right\rangle _{w}\right)\ell_{\alpha}\right)}.
\]
This agrees with the relation imposed by the Kubo-Martin-Schwinger
(KMS) condition on the equilibrium reservoir correlation functions
\cite{Kossakowski1977,Breuer2002}. Since throughout this paper we
consider only thermalized environments, we extend the above relation
to non-equilibrium situations
\begin{align}
\hat{\gamma}_{\alpha} & \left(\left\langle \mu_{c}\right\rangle _{w},\left\langle \mu_{v}\right\rangle _{w},\left\langle \psi\right\rangle _{w}\right)=\label{eq: quantum detailed balance rates}\\
 & =e^{-\beta\left(\hbar\omega_{\alpha}-\left(\left\langle \mu_{\alpha}\right\rangle _{w}+q\left\langle \psi\right\rangle _{w}\right)\ell_{\alpha}\right)}\gamma_{\alpha}\left(\left\langle \mu_{c}\right\rangle _{w},\left\langle \mu_{v}\right\rangle _{w},\left\langle \psi\right\rangle _{w}\right)\nonumber 
\end{align}
with $\mu_{\alpha\in I_{e}}=\mu_{c}$ and $\mu_{\alpha\in I_{h}}=\mu_{v}$.
For charge-conserving processes we require $\ell_{\alpha\in I_{0}}=0$,
single electron-capture processes are described by $\ell_{\alpha\in I_{e}}=-1$
and for single hole-capture processes it holds $\ell_{\alpha\in I_{h}}=+1$. 

Thus, supposing Eq.~(\ref{eq: quantum detailed balance rates}),
the hybrid model obeys the quantum detailed balance condition for
any model of the forward transition rate $\gamma_{\alpha}\left(\left\langle \mu_{c}\right\rangle _{w},\left\langle \mu_{v}\right\rangle _{w},\left\langle \psi\right\rangle _{w}\right)\geq0$
that is non-negative. Physically, the latter one must represent a
parametrization of a microscopically derived transition rate (using
Fermi's Golden Rule \cite{Alicki1977}) in terms of the averaged macroscopic
potentials. In particular, this enables the direct inclusion of microscopically
calculated capture rates e.g. from Refs.~\cite{Magnusdottir2002,Nielsen2004,Malic2007,Dachner2010,Wilms2013b}.

\subsection{Entropy production and the second law of thermodynamics\label{subsec:Entropy-production-and}}

From Eq.~(\ref{eq: entropy production (general)}) we obtain the
entropy production rate as (see Appendix~\ref{sec:Entropy-production-rate} for the derivation)
\begin{align}
\frac{\mathrm{d}S_{\text{tot}}}{\mathrm{d}t} & =\frac{1}{T}\int_{\Omega}\mathrm{d}^{3}r\,\left(\mu_{c}-\mu_{v}\right)R\nonumber \\
 & \phantom{=}+\frac{1}{qT}\int_{\Omega}\mathrm{d}^{3}r\,\left(\mathbf{j}_{n}\cdot\nabla\mu_{c}+\mathbf{j}_{p}\cdot\nabla\mu_{v}\right)\nonumber \\
 & \phantom{=}-k_{B}\mathrm{tr}\left(\left(\beta H+\log{\rho}\right)\mathcal{D}_{0}\left(\rho;\chi_w \right)\right)\label{eq: entropy production rate}\\
 & \phantom{=}-k_{B}\mathrm{tr}\left(\big(\beta\left(H-\mu_{c}^{\text{eff}}N\right)+\log{\rho}\big)\mathcal{D}_{e}\left(\rho;\chi_w\right)\right)\nonumber \\
 & \phantom{=}-k_{B}\mathrm{tr}\left(\big(\beta\left(H-\mu_{v}^{\text{eff}}N\right)+\log{\rho}\big)\mathcal{D}_{h}\left(\rho;\chi_w\right)\right)\nonumber 
\end{align}
with $\mu_{c/v}^{\text{eff}}=\left\langle \mu_{c/v}\right\rangle _{w}+q\left\langle \psi\right\rangle _{w}$. 
The dependency of the dissipators on the state of the classical environment is indicated by the abbreviation $\chi_w = \left( \langle \mu_c\rangle_w,\langle \mu_v\rangle_w,\langle \psi \rangle_w\right)$. 
The first two lines describe the entropy production rate of the van Roosbroeck
system \cite{Gajewski1996} and the third line is the entropy
production rate of an open quantum system coupled to a heat bath \cite{Spohn1978}.
The fourth and fifth line represent the contributions arising
from the coupling of the QD with its macroscopic environment
via capture and escape. All terms are products of abstract thermodynamic
forces and their corresponding fluxes, which is in agreement with
the general theory of linear irreversible thermodynamics \cite{DeGroot1984}.
Using \emph{Spohn's inequality} \cite{Spohn1978}, it can be shown that all individual lines of Eq.~(\ref{eq: entropy production rate})
are non-negative and therefore it holds
\[
\frac{\mathrm{d}S_{\text{tot}}}{\mathrm{d}t}\geq0,
\]
where the equality holds only in the case of thermodynamic equilibrium.
A proof is given in the Appendix~\ref{subsec:Positivity-of-the}.
This results relies on the specific coupling imposed in the previous
sections, which involves the spatially averaged macroscopic potentials.
We emphasize, that if e.g. averaged carrier densities were used instead,
a non-negative entropy production rate could not be guaranteed in
general. Finally, we conclude that our hybrid quantum-classical modeling
approach is consistent with the second law of thermodynamics.

Our approach can also be interpreted
as a damped Hamiltonian system in the framework of GENERIC \emph{(general
equation for the non-equilibrium reversible-irreversible coupling)}
\cite{Grmela1997}, which automatically ensures a non-negative
entropy production rate and the existence of an unique thermodynamic
equilibrium. It can be applied to a wide range of physical problems
\cite{Oettinger2011,Mielke2015,Mittnenzweig2017}. 

\section{Application to electrically driven single-photon sources\label{sec:Application}}

In this section we demonstrate the usefulness of our approach for applications in  semiconductor device simulation.
As an example we consider an electrically driven single-photon source
based on a p-i-n diode including a single QD. Such devices have been
shown to act as single-photon emitters and are promising candidates
for applications in quantum communication networks \cite{Yuan2002,Bennett2008,Unrau2012,Schlehahn2016a}.

\subsection{Model specification}

The model equations are described in Sec.~\ref{sec:Model-equations}
and \ref{sec:Thermodynamics}. For the hybrid system (\ref{eq: Poisson equation})--(\ref{eq: quantum master equation}),
we have to specify the Hamiltonian $H$ as well as the quantum jump
operators $A_{\alpha}$ and the transition rates $\gamma_{\alpha}$,
which constitute the dissipative interactions with the macroscopic
environment. In particular, they need to satisfy the conditions (\ref{eq: commutator N and H})
and (\ref{eq: charge conservation of D0}) that guarantee charge
conservation and the eigenoperator relations (\ref{eq: jump operator requirements}).

\begin{figure}[t]
\includegraphics[width=1\columnwidth]{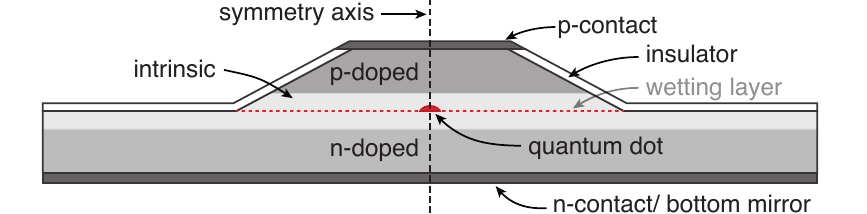}

\caption{Cross section of the example device considered in the numerical simulations:
A single QD is placed on the symmetry axis within the center of the
intrinsic zone of a cylindrical p-i-n diode with etched mesa structure
on top. The bottom mirror leads to a directed emission in vertical
direction. Due to the lack of a top mirror, the device represents
a leaky photonic cavity with low $Q$ factor. The device has electric
contacts at the top and the bottom facets.}

\label{fig: device sketch}
\end{figure}

\subsubsection{Hamiltonian}

We consider a single QD embedded in a very leaky dielectric cavity
with low $Q$ factor, which is sketched in Fig.~\ref{fig: device sketch}.
In such devices, the light-matter interaction is governed by spontaneous
emission and thus can be described by a Lindblad dissipator. Hence,
we can model the quantum system by a purely electronic Hamiltonian.
We aim for a description of the electronic QD states in terms of many-body
states covering single particle states, excitons, trions and the biexciton
as shown in Fig.~\ref{fig: electronic states}(a). We assume a single
one-particle level (ground state) for the electrons and holes each,
labeled by $\varepsilon_{c}$ and $\varepsilon_{v}$, respectively.
The Hamiltonian
\begin{align*}
H=H_{0}+H_{I}
\end{align*}
contains the single-particle contributions 
\begin{align*}
H_{0}=\sum_{\sigma}\varepsilon_{c}e_{\sigma}^{\dagger}e_{\sigma}-\sum_{\sigma}\varepsilon_{v}h_{\sigma}^{\dagger}h_{\sigma}
\end{align*}
and their Coulomb interaction
\begin{align*}
H_{I} & =\frac{1}{2}\sum_{\sigma,\sigma^{\prime}}\big(V_{c,c}e_{\sigma}^{\dagger}e_{\sigma^{\prime}}^{\dagger}e_{\sigma^{\prime}}e_{\sigma}+V_{v,v}h_{\sigma}^{\dagger}h_{\sigma^{\prime}}^{\dagger}h_{\sigma^{\prime}}h_{\sigma}-\\
 & \phantom{=\frac{1}{2}\sum\big(}-2V_{c,v}e_{\sigma}^{\dagger}h_{\sigma^{\prime}}^{\dagger}h_{\sigma^{\prime}}e_{\sigma}\big).
\end{align*}
The operators $e_{\sigma}^{\dagger}$ $(e_{\sigma})$ and $h_{\sigma}^{\dagger}$
$(h_{\sigma})$ create (annihilate) an electron or hole with total
angular momentum quantum number in $z$-direction $\sigma$. We consider
a single valence band describing heavy holes with a pseudo spin $\pm3/2$
indicated by $\left\{ \Uparrow,\Downarrow\right\} $. Here, only
Hartree-like Coulomb matrix elements $V_{i,j}=V_{i,j,j,i}$ occur,
which are of the order of several tens of meV (see Appendix \ref{sec:Parameters-and-auxiliary}).
The creation and annihilation operators obey the fermionic anti-commutator
relations $\lbrace e_{\sigma},e_{\sigma^{\prime}}^{\dagger}\rbrace=\lbrace h_{\sigma},h_{\sigma^{\prime}}^{\dagger}\rbrace=\delta_{\sigma,\sigma^{\prime}}$ and $\lbrace e_{\sigma},e_{\sigma^{\prime}}\rbrace=\lbrace h_{\sigma},h_{\sigma^{\prime}}\rbrace=0$. The single-particle
energy levels and the Coulomb matrix elements are obtained from Schr\"{o}dinger's
equation with an effective confinement potential for InGaAs-QDs \cite{Wojs1996}.
With the number operators $n_{e,\sigma}=e_{\sigma}^{\dagger}e_{\sigma}$,
$n_{h,\sigma}=h_{\sigma}^{\dagger}h_{\sigma}$ and the abbreviations
\begin{align*}
n_{e}=\sum_{\sigma=\left\{ \uparrow,\downarrow\right\} }n_{e,\sigma},\qquad n_{h}=\sum_{\sigma=\left\{ \Uparrow,\Downarrow\right\} }n_{h,\sigma} & ,
\end{align*}
we can express the Hamiltonian in the occupation number representation as
\begin{equation}
\begin{aligned}H & =\left(\varepsilon_{c}-\frac{1}{2}V_{c,c}\right)n_{e}-\left(\varepsilon_{v}+\frac{1}{2}V_{v,v}\right)n_{h}\\
 & \phantom{=}+\frac{1}{2}V_{c,c}n_{e}^{2}+\frac{1}{2}V_{v,v}n_{h}^{2}-V_{c,v}n_{e}n_{h}.
\end{aligned}
\label{eq: application Hamiltonian}
\end{equation}
By diagonalization, we obtain the spectral representation of $H$
in terms of multi-particle states
\[
H=\sum_{k}\varepsilon_{k}\big\vert k\big\rangle\big\langle k\big\vert,
\]
where $k=\left(n_{e,\uparrow},n_{e,\downarrow},n_{h,\Uparrow},n_{h,\Downarrow}\right)$
is a multi-index labeling the 16 different electronic configurations
which are illustrated in Fig.~\ref{fig: electronic states}(a, b).
 If excited states are included and full configuration interaction
is taken into account, the diagonalization of $H$ is in general a
non-trivial task. In this case, an approximative representation of
the Coulomb interaction in terms of number operators as in Eq.~(\ref{eq: application Hamiltonian})
can be obtained by the Hartree-Fock approximation \cite{Baer2004}.

\subsubsection{Dissipators}

\begin{figure}[t]
\includegraphics[width=1\columnwidth]{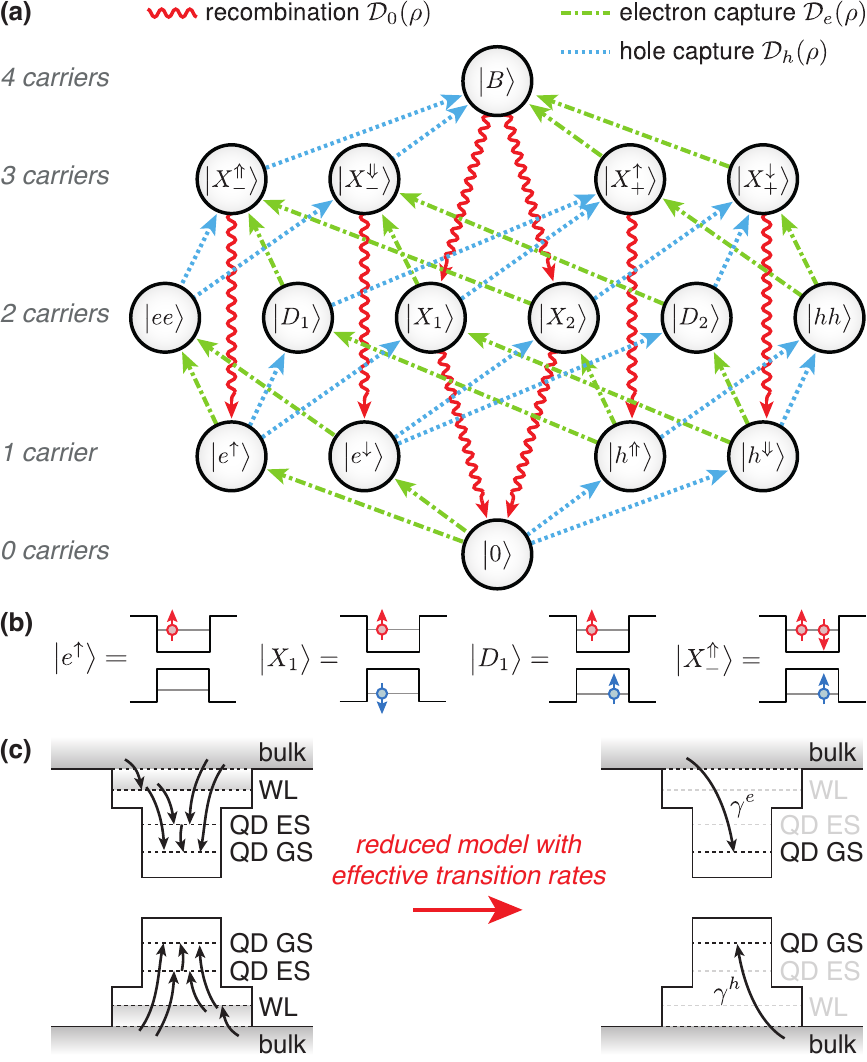}

\caption{(a)~Diagram of electronic states of the QD-Hamiltonian (\ref{eq: application Hamiltonian})
and possible (irreversible) transitions. The arrows indicate capture
and recombination, for the corresponding reverse processes (escape,
generation) the arrows need to be reversed. We use short notations
for the multi-particle states $\big\vert n_{e,\uparrow},n_{e,\downarrow},n_{h,\Uparrow},n_{h,\Downarrow}\big\rangle$:
empty QD $\big\vert0\big\rangle=\big\vert0,0,0,0\big\rangle$, single-electron
states $\big\vert e^{\uparrow}\big\rangle=\big\vert1,0,0,0\big\rangle$,
$\big\vert e^{\downarrow}\big\rangle=\big\vert0,1,0,0\big\rangle$,
single-hole states $\big\vert h^{\Uparrow}\big\rangle=\big\vert0,0,1,0\big\rangle$,
$\big\vert h^{\Downarrow}\big\rangle=\big\vert0,0,0,1\big\rangle$,
two-electron state $\big\vert ee\big\rangle=\big\vert1,1,0,0\big\rangle$,
two-hole state $\big\vert hh\big\rangle=\big\vert0,0,1,1\big\rangle$,
bright excitons $\big\vert X_{1}\big\rangle=\big\vert1,0,0,1\big\rangle$,
$\big\vert X_{2}\big\rangle=\big\vert0,1,1,0\big\rangle$, dark excitons
$\big\vert D_{1}\big\rangle=\big\vert1,0,1,0\big\rangle$, $\big\vert D_{2}\big\rangle=\big\vert0,1,0,1\big\rangle$,
negative trions $\big\vert X_{-}^{\Uparrow}\big\rangle=\big\vert1,1,1,0\big\rangle$,
$\big\vert X_{-}^{\Downarrow}\big\rangle=\big\vert1,1,0,1\big\rangle$,
positive trions $\big\vert X_{+}^{\uparrow}\big\rangle=\big\vert1,0,1,1\big\rangle$,
$\big\vert X_{+}^{\downarrow}\big\rangle=\big\vert0,1,1,1\big\rangle$
and the biexciton state $\big\vert B\big\rangle=\big\vert1,1,1,1\big\rangle$.
(b)~Schematic representation of the QD occupation for some example
states. (c)~Illustration of the effective scattering cascade in the
reduced model involving only the single-particle ground states.}

\label{fig: electronic states}
\end{figure}

We describe the spontaneous emission and the capture and escape of
carriers by dissipators of the type (\ref{eq: Dissipator (quantum detailed balance)}).
Even though the Hamiltonian of the quantum system Eq.~(\ref{eq: application Hamiltonian}) has a degenerate energy spectrum (due to spin degeneracy), the coherences are decoupled from the populations because of the selection rules. Hence, the resulting dynamical system reduces to a master equation for the populations \cite{Cuetara2016}. In this case, a jump operator $A_{\alpha}$ describes a transition between two
multi-particle states $\left|i\right\rangle $ and $\left|f\right\rangle $
is given by the projector $\left|f\right\rangle \left\langle i\right|$.
The allowed transitions are indicated by arrows in Fig.~\ref{fig: electronic states}(a),
e.g. the dissipator connected with $A_{\alpha}=\left|X_{1}\right\rangle \left\langle e^{\uparrow}\right|$
describes the capture of a hole into a QD occupied by a single electron
leading to the formation of the bright exciton $\left|X_{1}\right\rangle $.
By using adjacency matrices to encode the allowed transitions shown
in Fig.~\ref{fig: electronic states}(a), the dissipation superoperators for all
processes can be written in a compact form as\begin{subequations}\label{eq: application dissipators}
\begin{align}
\mathcal{D}_{e}\left(\rho;\chi_w\right) & =\sum_{i,f}\mathcal{A}_{i,f}^{e}\gamma_{i\to f}^{e}(\chi_w)\times \label{eq: application dissipator electron capture}\\
&\phantom{ =\sum_{i,f}}\times\left(L_{\left|f\right\rangle \left\langle i\right|}\left(\rho\right)+e^{-\beta\Delta\varepsilon_{i,f}^{e}(\chi_w)}L_{\left|i\right\rangle \left\langle f\right|}\left(\rho\right)\right),  \nonumber\\
\mathcal{D}_{h}\left(\rho;\chi_w\right) & =\sum_{i,f}\mathcal{A}_{i,f}^{h}\gamma_{i\to f}^{h}(\chi_w)\times \label{eq: application dissipator hole capture}\\
&\phantom{ =\sum_{i,f}}\times\left(L_{\left|f\right\rangle \left\langle i\right|}\left(\rho\right)+e^{-\beta\Delta\varepsilon_{i,f}^{h}(\chi_w)}L_{\left|i\right\rangle \left\langle f\right|}\left(\rho\right)\right),\nonumber \\
\mathcal{D}_{0}\left(\rho\right) & =\sum_{i,f}\mathcal{A}_{i,f}^{0}\gamma_{i\to f}^{0}\times \label{eq: application dissipator spontaneous emission}\\
&\phantom{ =\sum_{i,f}}\times\left(L_{\left|f\right\rangle \left\langle i\right|}\left(\rho\right)+e^{-\beta\Delta\varepsilon_{i,f}^{0}}L_{\left|i\right\rangle \left\langle f\right|}\left(\rho\right)\right), \nonumber
\end{align}
\end{subequations}
where the indices $i$ and $f$ run over all multi-particle eigenstates.
Again, the dependency of the dissipation superoperators on the state of the classical environment is indicated by the abbreviation $\chi_w = \left( \langle \mu_c\rangle_w,\langle \mu_v\rangle_w,\langle \psi \rangle_w\right)$. 
In accordance with Eq.~(\ref{eq: quantum detailed balance rates}),
the effective transition energies are given as
\begin{align*}
\Delta\varepsilon_{i,f}^{e}(\chi_w) & = \varepsilon_{i}-\varepsilon_{f}-q\left\langle \psi\right\rangle _{w}-\left\langle \mu_{c}\right\rangle _{w},\\
\Delta\varepsilon_{i,f}^{h}(\chi_w) & =\varepsilon_{i}-\varepsilon_{f}+q\left\langle \psi\right\rangle _{w}+\left\langle \mu_{v}\right\rangle _{w},\\
\Delta\varepsilon_{i,f}^{0} & =\varepsilon_{i}-\varepsilon_{f}
\end{align*}
and the adjacency matrix elements encoding Pauli blocking and the
optical selection rules (conservation of total angular momentum) read
\begin{align*}
\mathcal{A}_{i,f}^{e} & =\delta_{\left\langle i\right|n_{e}\left|i\right\rangle +1,\left\langle f\right|n_{e}\left|f\right\rangle }\prod_{\sigma=\left\{ \Uparrow,\Downarrow\right\} }\delta_{\left\langle i\right|n_{h,\sigma}\left|i\right\rangle ,\left\langle f\right|n_{h,\sigma}\left|f\right\rangle },\\
\mathcal{A}_{i,f}^{h} & =\delta_{\left\langle i\right|n_{h}\left|i\right\rangle +1,\left\langle f\right|n_{h}\left|f\right\rangle }\prod_{\sigma=\left\{ \uparrow,\downarrow\right\} }\delta_{\left\langle i\right|n_{e,\sigma}\left|i\right\rangle ,\left\langle f\right|n_{e,\sigma}\left|f\right\rangle },\\
\mathcal{A}_{i,f}^{0} & =\delta_{\left\langle i\right|n_{e,\uparrow}\left|i\right\rangle ,\left\langle f\right|n_{e,\uparrow}\left|f\right\rangle }\delta_{\left\langle i\right|n_{e,\downarrow}\left|i\right\rangle -1,\left\langle f\right|n_{e,\downarrow}\left|f\right\rangle }\times\\
 & \phantom{=+}\times\delta_{\left\langle i\right|n_{h,\Uparrow}\left|i\right\rangle -1,\left\langle f\right|n_{h,\Uparrow}\left|f\right\rangle }\delta_{\left\langle i\right|n_{h,\Downarrow}\left|i\right\rangle ,\left\langle f\right|n_{h,\Downarrow}\left|f\right\rangle }+\\
 & \phantom{=}+\delta_{\left\langle i\right|n_{e,\uparrow}\left|i\right\rangle -1,\left\langle f\right|n_{e,\uparrow}\left|f\right\rangle }\delta_{\left\langle i\right|n_{e,\downarrow}\left|i\right\rangle ,\left\langle f\right|n_{e,\downarrow}\left|f\right\rangle }\times\\
 & \phantom{=+}\times\delta_{\left\langle i\right|n_{h,\Uparrow}\left|i\right\rangle ,\left\langle f\right|n_{h,\Uparrow}\left|f\right\rangle }\delta_{\left\langle i\right|n_{h,\Downarrow}\left|i\right\rangle -1,\left\langle f\right|n_{h,\Downarrow}\left|f\right\rangle }.
\end{align*}
Please note that the above adjacency matrices are non-symmetric $\mathcal{A}_{i,f}\neq \mathcal{A}_{f,i}$. Thereby, they contain a \emph{directionality} which refers to the primary processes indicated by the arrow directions shown in Fig.~\ref{fig: electronic states}(a).
This is employed in the notation of the dissipators in Eq.~(\ref{eq: application dissipators}), 
which explicitly accounts for the actual net-transition rates by hard-wiring the relation (\ref{eq: quantum detailed balance rates}) between forward and backward rates. Consequently, the quantum detailed balance relation is guaranteed in the thermodynamic equilibrium independent of the model for the forward rate $\gamma_{i\to f}(\chi_w)$. The respective backward transition rate is obtained according to Eq.~(\ref{eq: quantum detailed balance rates}).
 An alternative representation of the master equation for the populations can be found in Appendix~\ref{sec:Projection-on-eigenstates}.

\subsubsection{Transition rate models}

The spontaneous decay rates of the various (bright) electronic states
of the quantum system can be modeled by the Weisskopf-Wigner rate \cite{Weisskopf1930}
\begin{align}
\gamma_{i\to f}^{0} & =\frac{P_{i,f}d_{c,v}^{2}n_{r}}{6\pi\hbar\varepsilon_{0}c_{0}^{3}}\left(\frac{\varepsilon_{i}-\varepsilon_{f}}{\hbar}\right)^{3}\left(1+n_{\text{pt}}\left(\frac{\varepsilon_{i}-\varepsilon_{f}}{\hbar}\right)\right)\label{eq: radiative decay rate}
\end{align}
for allowed index pairs ${i\to f}$ giving ${\mathcal{A}_{i,f}^0 = 1}$.
Here, $n_{\text{pt}}\left(\omega\right)=\left(e^{\beta\hbar\omega}-1\right)^{-1}$
is the thermally induced photon number, $n_{r}$ is the refractive
index of the material, $d_{c,v}$ denotes the interband dipole moment
and $c_{0}$ is the vacuum speed of light. Due to cavity effects,
the decay rate is slightly modified with respect to the free space
decay rate, which is accounted for by the Purcell factors $P_{i,f}$.
The Weisskopf-Wigner rate is applicable in low $Q$ optical resonators,
where the photonic density of states varies insignificantly over the
linewidth of the emitter \cite{Lodahl2015,Florian2012}. Using the
parameters given in Appendix~\ref{sec:Parameters-and-auxiliary},
all decay rates are found to be approximately $10^{9}\,\text{s}^{-1}$.

For semiconductor QDs, the Fr\"{o}hlich coupling and Auger scattering typically
constitute the dominant capture processes. As a rule of thumb,
at low carrier densities, the LO-phonon assisted
Fr\"{o}hlich coupling provides the dominant scattering channel, whereas
at elevated carrier densities the Auger scattering becomes
increasingly efficient \cite{Dachner2010,Chow2013,Ferreira2015}.
Due to the relatively large Coulomb matrix elements in semiconductor
QDs, the scattering rates into charged states differ significantly
from those into neutral states. This effect is known as \emph{Coulomb
suppression} or \emph{Coulomb enhancement}, respectively \cite{Ferreira2015}.

The scattering rates can be calculated microscopically by Fermi's
Golden rule and then always satisfy the detailed balance relation
between the forward and backward process \cite{Magnusdottir2002,Nielsen2004,Malic2007,Dachner2010,Wilms2013b}.
However, here we restrict ourselves to phenomenological laws for the effective capture rate of
continuum carriers into the QD. The respective escape rates follow via the detailed balance relation.
The effective capture rate approximates the entire scattering cascade, see Fig.~\ref{fig: electronic states}(c).
We model the effective electron capture rates entering Eq.~(\ref{eq: application dissipator electron capture})
as
\begin{subequations}
\label{eq: capture rate model}
\begin{align}
\gamma_{i\to f}^{e} & \left(\left\langle \mu_{c}\right\rangle _{w},\left\langle \mu_{v}\right\rangle _{w},\left\langle \psi\right\rangle _{w}\right)=\label{eq: capture rate model electrons}\\
 & =\frac{1+n_{\text{LO}}}{\tau_{\text{LO}}^{e}}\frac{1}{e^{\beta\left(E_{c}-q\left\langle \psi\right\rangle _{w}-\left\langle \mu_{c}\right\rangle _{w}+a_{\text{LO}}^{e}+C_{i,f}^{e}\right)}+1}+\nonumber \\
 & \phantom{=}+\frac{1}{\tau_{\text{Au}}^{e,e}}\frac{\bar{n}_{w}^{2}}{1+\bar{n}_{w}^{2-\gamma_{\text{Au}}^{e,e}}}+\frac{1}{\tau_{\text{Au}}^{e,h}}\frac{\bar{n}_{w}\bar{p}_{w}}{1+\left(\bar{n}_{w}\bar{p}_{w}\right)^{1-\gamma_{\text{Au}}^{e,h}/2}},\nonumber 
\end{align}
for the admitted index pairs ${i\to f}$ giving ${\mathcal{A}_{i,f}^e = 1}$
and the effective hole capture rates in Eq. (\ref{eq: application dissipator hole capture})
as
\begin{align}
\gamma_{i\to f}^{h} & \left(\left\langle \mu_{c}\right\rangle _{w},\left\langle \mu_{v}\right\rangle _{w},\left\langle \psi\right\rangle _{w}\right)=\label{eq: capture rate model holes}\\
 & =\frac{1+n_{\text{LO}}}{\tau_{\text{LO}}^{h}}\frac{1}{e^{-\beta\left(E_{v}-q\left\langle \psi\right\rangle _{w}-\left\langle \mu_{v}\right\rangle -a_{\text{LO}}^{h}-C_{i,f}^{h}\right)}+1}+\nonumber \\
 & \phantom{=}+\frac{1}{\tau_{\text{Au}}^{h,h}}\frac{\bar{p}_{w}^{2}}{1+\bar{p}_{w}^{2-\gamma_{\text{Au}}^{h,h}}}+\frac{1}{\tau_{\text{Au}}^{h,e}}\frac{\bar{n}_{w}\bar{p}_{w}}{1+\left(\bar{n}_{w}\bar{p}_{w}\right)^{1-\gamma_{\text{Au}}^{h,e}/2}}.\nonumber 
\end{align}
\end{subequations}
for the admitted index pairs ${i\to f}$ yielding ${\mathcal{A}_{i,f}^h = 1}$.
Here we have used the abbreviations $\bar{n}_{w}=n_{w}/n_{\text{Au}}^{\text{crit}}$
and $\bar{p}_{w}=p_{w}/p_{\text{Au}}^{\text{crit}}$. Please note
that the ambient continuum carrier densities ${n_{w}=N_{c}F_{1/2}\left(\beta\left(\left\langle \mu_{c}\right\rangle _{w}-E_{c}+q\left\langle \psi\right\rangle _{w}\right)\right)}$
and ${p_{w}=N_{v}F_{1/2}\left(\beta\left(E_{v}-q\left\langle \psi\right\rangle _{w}-\left\langle \mu_{v}\right\rangle _{w}\right)\right)}$
are functions of the averaged macroscopic potentials. The first terms
in Eq.~(\ref{eq: capture rate model}) describe the LO-phonon assisted relaxation of continuum carriers and
the last lines are each attributed to Auger scattering. The number
of thermally excited LO-phonons is given by $n_{\text{LO}}=\left(e^{\beta\hbar\omega_{\text{LO}}}-1\right)^{-1}$.
The time constants $\tau_{\text{LO}}^{\lambda}$
and the parameters $a_{\text{LO}}^{\lambda},\gamma_{\text{LO}}^{\lambda}$,
$\lambda\in\left\{ e,h\right\} $ are considered
as fitting factors that can be extracted from microscopic calculations
or experimental data. The phonon assisted capture rates involve the
Coulomb enhancement/\,suppression factors
\begin{align*}
C_{i,f}^{e} & =\varepsilon_{f}-\varepsilon_{i}-\varepsilon_{c},\\
C_{i,f}^{h} & =\varepsilon_{f}-\varepsilon_{i}+\varepsilon_{v},
\end{align*}
which describe the additional attractive or repulsive Coulomb shifts
and thereby either enhance (if $C_{i,f}^{\lambda}<0$, $\lambda\in\left\{ e,h\right\} $)
or decrease (if $C_{i,f}^{\lambda}>0$, $\lambda\in\left\{ e,h\right\} $)
the capture rate. At low temperatures the effect of Coulomb enhancement
or suppression becomes increasingly important. For the Auger-like
capture processes the modifications of the capture rates due to Coulomb
shifts are assumed to be negligible due to strong screening effects
at high carrier densities. The expressions in Eq.~(\ref{eq: capture rate model}) take saturation effects
at high carrier densities into account. The functional form
is motivated from microscopically computed results presented in Refs.~\cite{Wilms2013b,Ferreira2015}.
In the low density limit (Maxwell--Boltzmann approximation)
the capture rate models asymptotically take the form
\begin{align*}
\gamma_{i\to f}^{e} & \approx\frac{\left(n_{\text{LO}}+1\right)e^{-\beta a_{\text{LO}}^{e}}}{\tau_{\text{LO}}^{e}N_{c}}e^{-\beta C_{i,f}^{e}}n_{w}^{\text{MB}}\\
 & \phantom{=}+\frac{\left(\bar{n}_{w}^{\text{MB}}\right)^{2}}{\tau_{\text{Au}}^{e,e}}+\frac{\bar{n}_{w}^{\text{MB}}\bar{p}_{w}^{\text{MB}}}{\tau_{\text{Au}}^{e,h}},\\
\gamma_{i\to f}^{h} & \approx\frac{\left(n_{\text{LO}}+1\right)e^{-\beta a_{\text{LO}}^{h}}}{\tau_{\text{LO}}^{h}N_{v}}e^{-\beta C_{i,f}^{h}}p_{w}^{\text{MB}}\\
 & \phantom{=}+\frac{\left(\bar{p}_{w}^{\text{MB}}\right)^{2}}{\tau_{\text{Au}}^{h,h}}+\frac{\bar{n}_{w}^{\text{MB}}\bar{p}_{w}^{\text{MB}}}{\tau_{\text{Au}}^{h,e}},
\end{align*}
showing a linear dependency on the continuum carrier density in the
case of LO-phonon assisted capture and a quadratic dependency for
the Auger capture processes. Moreover, the Coulomb enhancement and
suppression effect becomes apparent in this form. The expression for
$n_{w}^{\text{MB}}$ is obtained by replacing $F_{1/2}\left(\cdot\right)\to\exp{\left(\cdot\right)}$
in the above definition of $n_{w}$ (analogous for $p_{w}^{\text{MB}}$).
The parameters $n_{\text{Au}}^{\text{crit}}$, $p_{\text{Au}}^{\text{crit}}$
and $\gamma_{\text{Au}}^{\lambda,\lambda^{\prime}}$ $\lambda,\lambda^{\prime}\in\left\{ e,h\right\} $
are fitting factors.

\subsection{Numerical simulation method}

The van Roosbroeck system (\ref{eq: Poisson equation})--(\ref{eq: hole transport})
is discretized using a Vorono\"{i} box based finite volumes method
\cite{Selberherr1984,Farrell2017} along with a modified Scharfetter--Gummel
scheme \cite{Scharfetter1969,Bessemoulin-Chatard2012,Koprucki2014}
for the discretization of the current densities. The latter one properly
reflects the strong degeneration effects of the electron-hole plasma
at cryogenic temperatures and takes the Fermi--Dirac statistics
and nonlinear diffusion via a generalized Einstein relation fully
into account \cite{Kantner2016}. For time-dependent simulations,
we use an implicit Euler discretization and an adaptive time
stepping method.

The discretized van Roosbroeck system is solved along with the QME
(\ref{eq: quantum master equation}) by a full Newton iteration using
the electrostatic potential $\psi$, the quasi-Fermi energies $\mu_{c}$,
$\mu_{v}$ and the density matrix elements $\langle k\vert\rho\vert l\rangle$
as independent variables. In order to obtain a system of
ordinary differential equations, the QME is projected on the Hilbert
space basis spanned by the multi-particle eigenstates of $H$ (see
Appendix~\ref{sec:Projection-on-eigenstates}).

The coupling terms $Q$ and $S_{n/p}$ given by Eq.~(\ref{eq: charge density of the quantum system}) and (\ref{eq: loss terms}) introduce a non-local
coupling of the van Roosbroeck system with the QME via the spatial
profile function $w$. This has an impact on the sparsity pattern
of the Jacobian of the discretized system, since the quantum system
interacts in general with a large number of control volumes in its
environment. Since the discretized spatial profile function $w_{K}=\left|\Omega_{K}\right|^{-1}\int_{\Omega_{K}}\mathrm{d}^{3}r\,w\left(\mathbf{r}\right)$
(where $\left|\Omega_{K}\right|$ is the volume of the $K$-th Vorono\"{i}
cell), quickly decays, we discard small matrix elements below a chosen
threshold. This preserves the quadratic convergence of  Newton's
iteration while the numerical effort is reduced.

Single-photon sources are typically operated at cryogenic temperatures,
which causes serious convergence issues during the numerical solution
of the van Roosbroeck system because of the strong depletion of minority
carrier densities \cite{Selberherr1987,Richey1994,Kantner2016}. By using the
temperature embedding method described in Ref.~\cite{Kantner2016},
the problem becomes tractable in the vicinity of flat band conditions.

\subsection{Device specification}

In the numerical simulations presented in the following, we consider
the cylindrical GaAs-based p-i-n structure depicted in Fig.~\ref{fig: device sketch},
where a single QD is placed on the symmetry axis within the center
of the intrinsic zone. The total height of the device is $800\,\text{nm}$, the intrinsic layer has a thickness of $200\,\text{nm}$ and
the doped layers both are $300\,\text{nm}$ in height. The doping
concentrations are $C=N_{D}=2\times10^{18}\,\text{cm}^{-3}$ and $C=-N_{A}=-10^{19}\,\text{cm}^{-3}$
in the n- and p-domain, respectively. The top radius of the mesa is
0.5\,\textmu m and the total radius (at the bottom) is 2.5\,\textmu m.
The bottom facet is assumed to consist of a highly reflective metal
such that it simultaneously acts as an electric contact and a mirror
leading to a directed emission in vertical direction. The ohmic contact
on the top facet is assumed to consist of an optically transparent
material, such that the structure forms a leaky cavity with a low
$Q$ factor. The remaining facets are modeled by homogeneous Neumann
boundary conditions. The wetting layer (WL) indicated in Fig.~\ref{fig: device sketch}
is neglected in the simulation. The device is assumed to operate under
cryogenic conditions at $T=50\,\text{K}$.

The numerical simulation exploits the rotational symmetry of the device,
such that the computational domain reduces to a 2D cross section with
adapted cell volumes.

\subsection{Stationary operation}

\begin{figure}
\includegraphics[width=1\columnwidth]{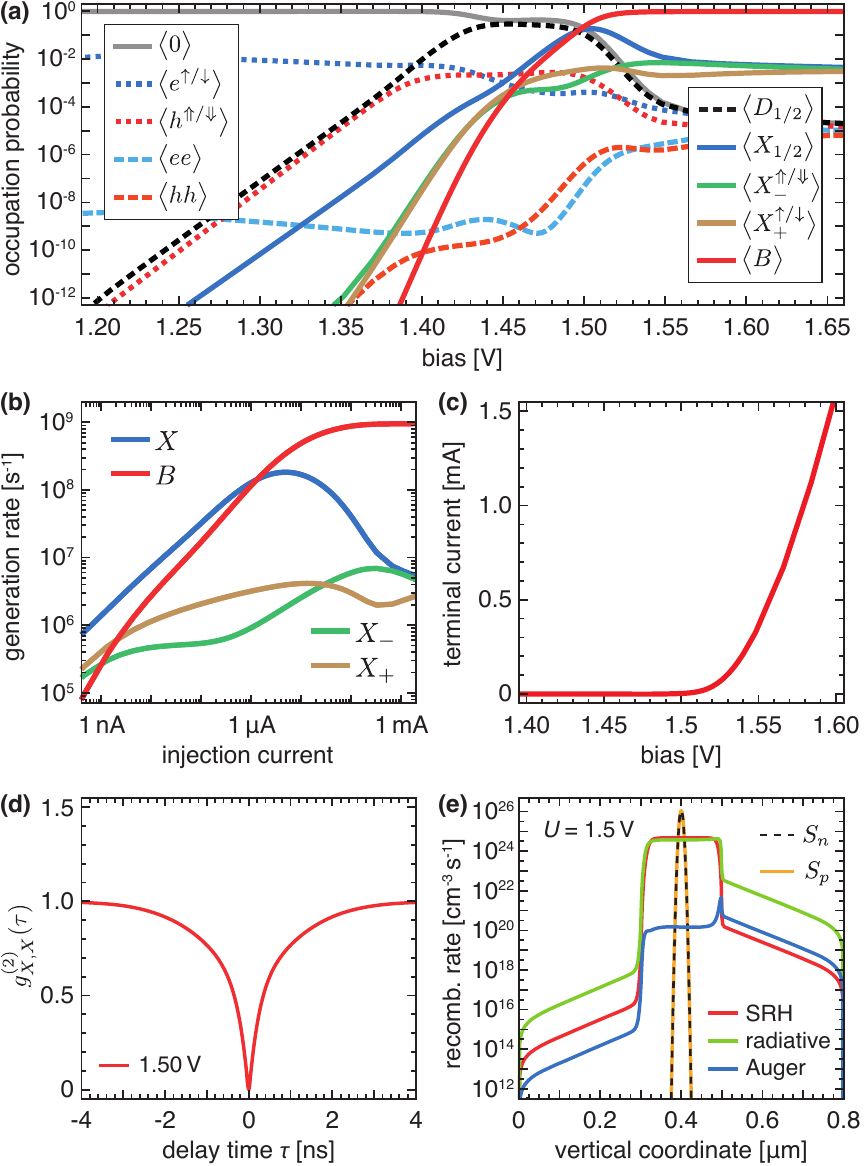}
\caption{Numerical results at stationary injection. (a)~Occupation of the QD
states vs. applied bias. (b)~Single-photon generation rates of the
different emission lines vs. injection current. (c)~Current-voltage
curve of the diode. (d)~Second order correlation function of the photons
generated on the exciton line. (e)~Comparison of recombination and
capture rates of free carriers along the symmetry axis of the device.
For the labeling of QD states we refer to the caption of Fig.~\ref{fig: electronic states}.}

\label{fig: stationary injection}
\end{figure}

The device operates as a p-i-n diode, which
can be seen from the current-voltage curve shown in Fig.~\ref{fig: stationary injection}(c).
At cryogenic temperatures the Fermi energy levels in the doped domains
are very close to the band edges and therefore the diode's threshold
voltage approximately equals the energy band gap of the material (around
1.52\,V). The population of the QD states $\langle k\rangle =\langle k\vert\rho\vert k\rangle$
can be controlled by the externally applied bias as shown in Fig.~\ref{fig: stationary injection}(a).
Since the QD is located within the intrinsic zone of the device,
it is most probably unoccupied in the low bias regime. When the applied
bias approaches the diode's threshold voltage, the QD population turns
into a non-equilibrium distribution: At first, due to the increased
continuum carrier densities in the vicinity of the QD, the single-particle
and excitonic states are populated. In particular, due to the lack
of an radiative decay channel, the dark excitons $\langle D_{1/2}\rangle$
have a high occupation probability. Finally, beyond the threshold,
the QD is quickly driven into saturation and the population is dominated
by the biexciton state $\langle B\rangle$. Due to Coulomb
enhancement and suppression, the population of neutral states is favored
in the whole bias range. In particular, Fig.~\ref{fig: stationary injection}(a)
shows that the population of the doubly charged states $\langle ee\rangle$
and $\langle hh\rangle$ is strongly suppressed.

The single-photon generation rates of the different emission lines
are given by
\begin{equation}
\Gamma_{k}=\sum_{l}\mathcal{A}_{k,l}^{0}\gamma_{k \to l }^{0}\langle k\rangle .\label{eq: single photon generation rate}
\end{equation}
Since the decay rates for all radiative processes are approximately
equal, the occupation probabilities are directly proportional to the
single-photon generation rates, which are depicted in Fig.~\ref{fig: stationary injection}(b).
At low injection currents, the emission spectrum is dominated by photons
generated via the decay of bright excitons. Close to the threshold
voltage the bright exciton line reaches a maximum and then decreases
while the intensity of the biexciton line grows until it finally saturates.
In this regime, the capture rates exceed the radiative decay rates
by several orders of magnitude. This simulation result agrees with
experimental observations presented in Ref. \cite{Yuan2002}.

\begin{figure}
\includegraphics[width=1\columnwidth]{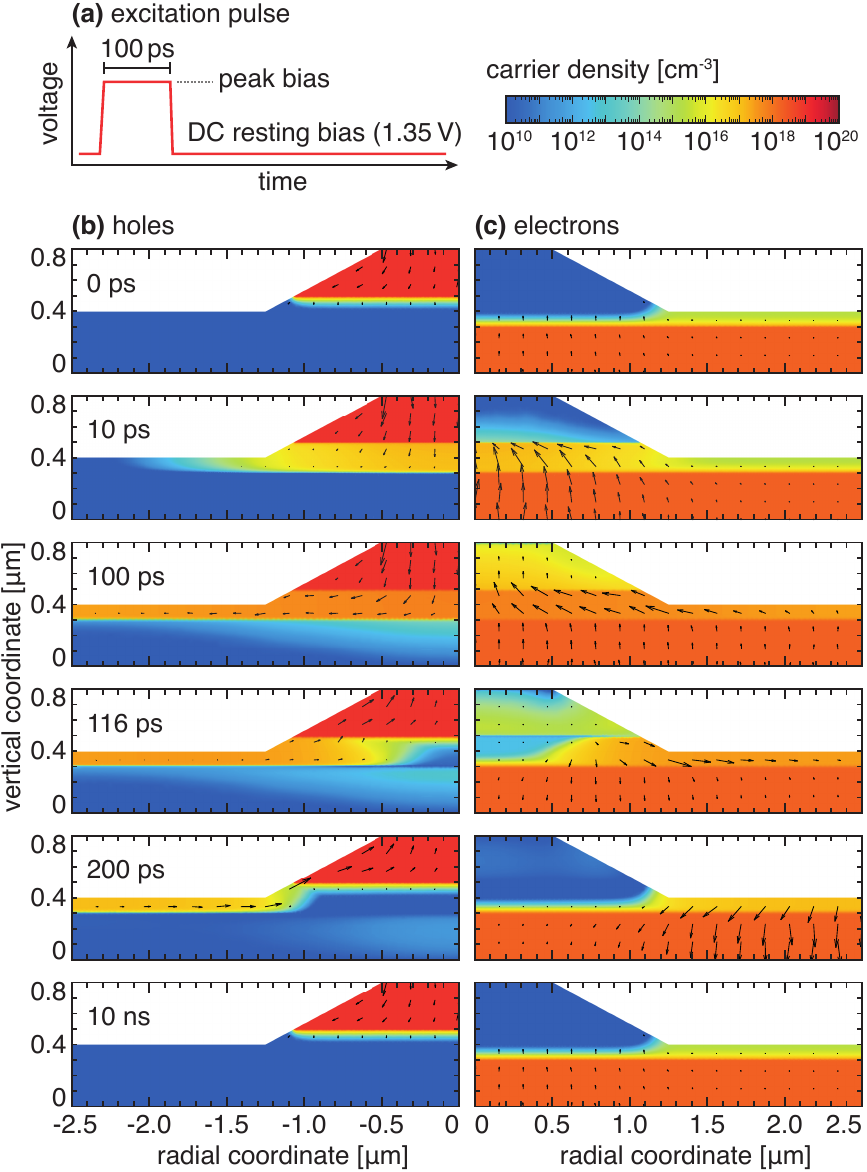}
\caption{Carrier transport at pulsed excitation. (a)~Illustration of the voltage
ramp used in the simulations. (b, c)~Snapshots of the carrier density
distribution on a 2D cross-section at several instances of time. The carrier density
is color-coded, the arrows indicate the current density vector field
(arrows point into the direction of particle motion). The peak voltage
in the simulation was set to $1.6\,\text{V}$.}

\label{fig: pulsed injection transport}
\end{figure}

Another important figure of merit for single-photon emitters is the
second order intensity correlation function of the generated photons
\begin{equation}
g^{\left(2\right)}\left(\tau\right)=\frac{\left\langle a^{\dagger}\left(0\right)a^{\dagger}\left(\tau\right)a\left(\tau\right)a\left(0\right)\right\rangle }{\left\langle a^{\dagger}\left(0\right)a\left(0\right)\right\rangle ^{2}},\label{eq: second order correlation function}
\end{equation}
where the operator $a^{\dagger}\left(a\right)$ creates (annihilates)
a photon and $\tau$ is a time delay. A value of $g^{\left(2\right)}\left(0\right)<0.5$
indicates the presence of a single-photon Fock state in the radiation
field. In our model the decay of an optically active QD state is equivalent
to the generation of a corresponding photon. Therefore, the electronic
operators can be used to evaluate Eq.~(\ref{eq: second order correlation function}),
cf. Ref.~\cite{Florian2012}. For the bright exciton line, we identify
the photon creation operator with the projector $a^{\dagger}=\vert0\rangle\langle X_{i}\vert$
(with $i=1$ or 2) and use the quantum regression theorem \cite{Breuer2002}
to evaluate Eq.~(\ref{eq: second order correlation function}). The
result is presented in Fig.~\ref{fig: stationary injection}(d) and
recovers the characteristic dip around $\tau = 0$ for high-quality
single-photon sources \cite{Yuan2002,Santori2010}. Since the present
model assumes an ideal quantum emitter and an instantaneous extraction
of the generated photons from the cavity, the value of $g^{\left(2\right)}\left(0\right)$
is exactly zero. For a refined description at this stage, a
coherent light-matter interaction must be included in the Hamiltonian
and $\mathcal{D}_{0}$ has to be extended by a photon
outcoupling mechanism.

Finally, in Fig.~\ref{fig: stationary injection}(e) we show the
recombination rate $R$  of the continuum
carriers and the capture rates $S_{n/p}$ along the vertical (symmetry) axis of the device. Close to
the threshold voltage, the transition of carriers into the QD imposes
the dominant loss mechanism of continuum carriers in the vicinity
of the QD.

\begin{figure}
\includegraphics[width=1\columnwidth]{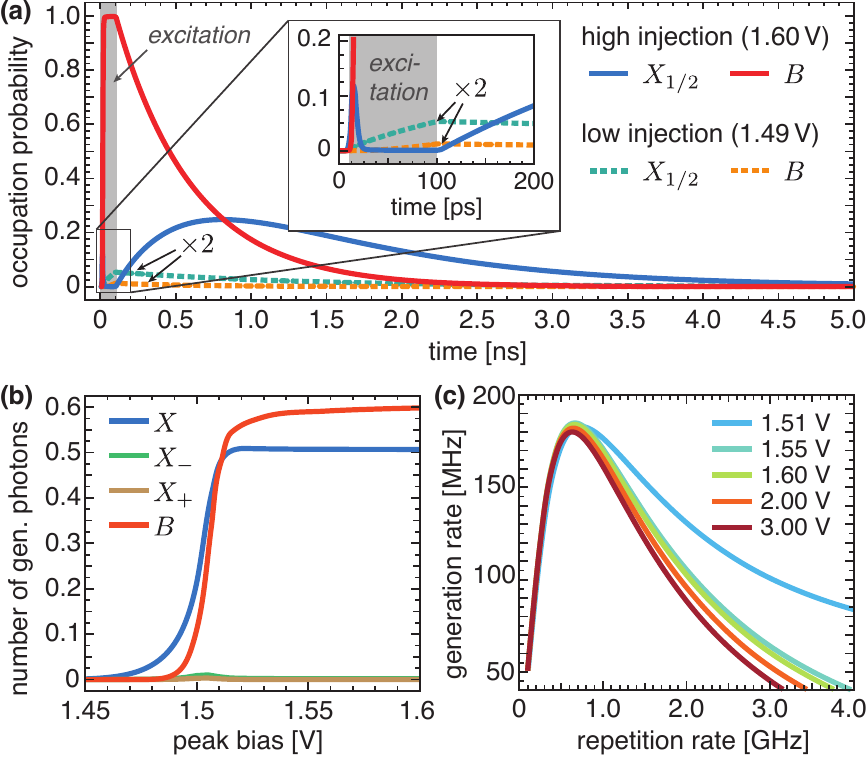}
\caption{QD occupation and single-photon generation under pulsed excitation.
(a)~Comparison of the transient exciton and biexciton occupation
probabilities in the low and high injection case. In the low injection
case the occupation probabilities have been multiplied by a factor
2 for better visibility. (b)~Number of generated photons per pulse
on the different emission lines for different values of the peak bias.
(c)~Single-photon generation rate on the bright exciton line vs.
repetition frequency of the time-periodic excitation pulse (for different
values of the peak bias).}

\label{fig: pulsed injection quantum dot}
\end{figure}
\subsection{Pulsed operation}

For many applications, the generation of single photons at certain
instances of time is required. Electrically driven QD-based single-photon
sources offer an easy off-resonant excitation scheme \cite{Michler2009, Florian2012},
where the QD is excited by short voltage pulses. This process shall
be simulated in the following, where we apply rectangular voltage
pulses with a fixed duration of 100\,ps superimposed on a DC bias
of 1.35\,V as illustrated in Fig.~\ref{fig: pulsed injection transport}(a).
We investigate the impact of the pulse repetition time and the peak
bias, which are the key external control parameters. The results of
a numerical carrier transport simulation for a single pulse with a
peak voltage of $1.6\,\text{V}$ are shown in Fig.~\ref{fig: pulsed injection transport}(b,
c). Due to the high carrier mobilities at low temperatures (cf. Appendix~\ref{sec:Parameters-and-auxiliary}),
the carriers quickly spread out within the device such that the intrinsic
zone is highly populated at the end of the excitation pulse (100\,ps).
Subsequently, when the applied voltage is switched back to the resting
DC bias, the carriers are quickly withdrawn from the intrinsic zone.
In the snapshots taken at $116\,\text{ps}$ and $200\,\text{ps}$
we observe that in particular the vicinity of the QD (which is located
on the center of the symmetry axis at 0.4\,\textmu m, cf. Fig.~\ref{fig: device sketch})
is depleted first. Moreover, a conducting channel underneath the insulating
region is formed. The plot at 10\,ns shows the stationary state reached
after a long time.

The impact of the voltage pulse on the occupation of the QD is shown
in Fig.~\ref{fig: pulsed injection quantum dot}(a). In the case
of an excitation with a peak voltage of $1.6\,\text{V}$ (high injection),
one first observes a fast occupation of the biexciton state which
subsequently decays radiatively. Via the so-called \emph{biexciton-cascade},
the bright exciton states are populated in the following. Comparing
the time scales of the carrier transport with the life times of the
bright QD states, see Fig.~\ref{fig: pulsed injection transport}(b,
c) and Fig.~\ref{fig: pulsed injection quantum dot}(a), it is apparent
that the decay of the bright exciton happens a long time after the
continuum carriers have left the vicinity of the QD. This separation
of time scales is of particular importance for the generation of indistinguishable
photons \cite{Bennett2008}, since fluctuations of the carrier density
in the vicinity of the emitter might shift the generated photon's
energy.

Next, we study the impact of the peak bias value. Figure~\ref{fig: pulsed injection quantum dot}(b)
shows the number of generated photons for different peak voltages
after 10\,ns. The number of generated photons on line $k$ until
time $t$ is obtained from 
\[
N_{k}\left(t\right)=\int_{0}^{t}\mathrm{d}t^{\prime}\,\Gamma_{k}\left(t^{\prime}\right),
\]
using the single-photon generation rate defined in Eq.~(\ref{eq: single photon generation rate}).
The plot clearly reveals the existence of two regimes: A subthreshold
(low injection) regime, where the peak voltage is insufficient for
the excitation of the QD (cf. Fig.~\ref{fig: pulsed injection quantum dot}(a)),
and a high injection regime where the biexciton-cascade can be observed
practically after each pulse. For the exciton-photons, this implies
a generation efficiency of around 50\,\% for both polarizations.
The generation efficiency of the two differently polarized photons
on the biexciton-line is a little higher than 50\,\%, due to additional
recombination during the excitation period, see Fig.~\ref{fig: pulsed injection quantum dot}(a,
b).

\begin{figure}
\includegraphics[width=1\columnwidth]{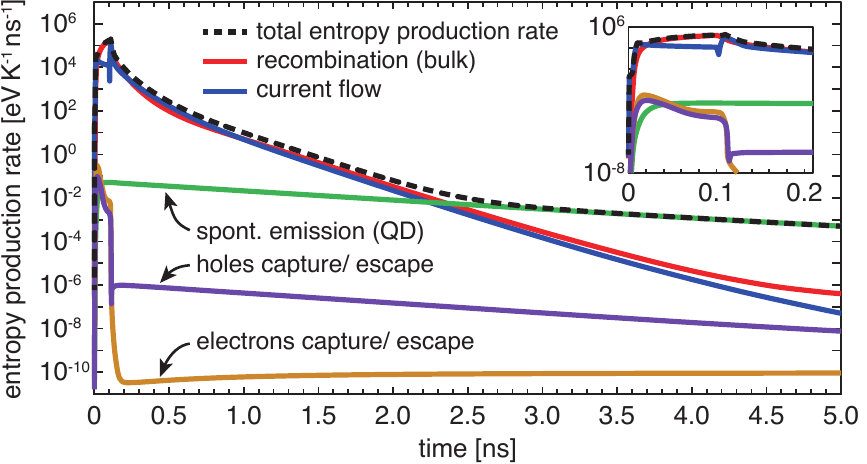}
\caption{Entropy production rate during pulsed excitation (peak bias 1.6\,V). The
plot shows the five contributions arising from the individual lines
in Eq.~(\ref{eq: entropy production rate}) and the total entropy
production rate. The inset is a zoom on the
first 200\,ps.}

\label{fig: pulsed injection entropy production}
\end{figure}

Finally, we investigate the optimal repetition frequency of the excitation
cycle for the generation of single exciton-photons. The optimal repetition
frequency $f^{\ast}=1/t^{\ast}$ maximizes the number of generated
photons per time: 
\[
\bar{\Gamma}_{X}\left(t^{\ast}\right)=\frac{N_{X}\left(t^{\ast}\right)}{t^{\ast}}=\frac{1}{t^{\ast}}\int_{0}^{t^{\ast}}\mathrm{d}t^{\prime}\,\Gamma_{X}\left(t^{\prime}\right)\to\max.
\]
Figure~\ref{fig: pulsed injection quantum dot}(c) shows a clear
maximum at a pulse repetition rate of $f^{\ast}\approx650\,\text{MHz}$
($t^{\ast}\approx1.5\,\text{ns}$), which corresponds to a maximum
single-photon generation rate of $\bar{\Gamma}_{X}\left(t^{\ast}\right)\approx185\,\text{MHz}$.
Even though in this optimal case the photon generation efficiency
per pulse shrinks to 28\%, the high repetition frequency leads to
an enhanced overall performance. Moreover, Fig.~\ref{fig: pulsed injection quantum dot}(c)
indicates that this result is practically independent of the peak
voltage. In order to obtain the actual emission rate, the generation
rate must be multiplied with the extraction efficiency \cite{Zwiller2004}.

We conclude this section with a consideration of the entropy production
during an excitation cycle, which is depicted in Fig.~\ref{fig: pulsed injection entropy production}.
The plot shows, that during the first 2\,ns the entropy production
rate is clearly governed by the contributions arising from the macroscopic
system, whereas at later times the slow decay of the QD-exciton becomes
dominant. The numerical result is in agreement with the theory presented
in Sec.~\ref{subsec:Entropy-production-and}, which predicts a positive
entropy production rate at all times.

\section{Discussion and outlook\label{sec:Outlook}}

The electrically driven single-photon source considered in the previous section is a realistic application that fits into the framework of the new model system (\ref{eq: Poisson equation})--(\ref{eq: quantum master equation}) introduced in this paper. This example is a \emph{proof of concept} that shows the computational tractability of our approach regarding its potential for applications in numerical semiconductor device simulation.
As demonstrated in Sec.~\ref{sec:Application}, the hybrid approach enables a comprehensive analysis of QD-based devices
even in the case of complex, multi-dimensional device geometries as well as the investigation of transient processes.

If the feedback of the quantum system on its classical environment is weak, 
i.e. if the capture rates $S_{n/p}$ are small compared to the recombination rate $R$, 
simplified approaches can be considered.
A first option is to merely consider the semi-classical transport while neglecting the quantum system as carried out e.g. in Ref.~\cite{Kantner2016a}. In this case, however, the model gives no access on the quantum optical figures of merit, of course.
A second option is to treat the quantum master equation alone by choosing an appropriate parametrization of the transition rates in the dissipation superoperators as done e.g. in Ref.~\cite{Florian2012}.
However, in electrically driven devices, the carrier densities, the electric field and the current densities, which usually drive the transition rates, strongly depend on the applied voltage and can vary over many orders of magnitude. In general, their detailed behavior is not apriori known and requires full device simulation since the evolution of these quantities is determined by specific design parameters such as the device geometry, doping profiles, heterostructures etc.
In conclusion, the hybrid modeling approach described in this paper goes beyond existing ones.

In the case of weak feedback, the coupling of both subsystems in the hybrid model becomes effectively uni-directional.
This means that the dynamics of the quantum system is slaved by the evolution of its classical environment, which can be exploited to reduce the computational effort in a two-step method:
First, the transport simulation is carried out whilst omitting coupling terms to the quantum system. 
In a second step, the solution of the classical system is used to determine the time-dependent dissipators that drive the evolution of the open quantum system.
Hence, the quantum master equation is solved in a ``post-processing'' step, 
which finally gives access to the quantum optical figures of merit.
Via the explicit dependency of the microscopic transition rates on the state of the classical environment (spatially averaged macroscopic potentials), the hybrid model provides a consistent link between the two steps of the unidirectionally coupled simulation approach. 
Nevertheless, even in the case of weak feedback, where one-way coupled approaches are admissible,
the fully coupled hybrid model allows to assess the approximation errors. Thereby it helps to justify simplified simulation approaches.

The application considered in Sec.~\ref{sec:Application} is an example for a quantum system with a weak feedback on the classical environment, which in principle would allow for the one-way method outlined above.
The essential reason for this is the slow radiative decay in comparison to the fast electronic processes, which keeps the capture rates $S_{n/p}$ small once the QD is occupied. However, this is not always the case. For example, in electrically driven QD nanolasers, where the QD is placed inside a resonant cavity, the Purcell-enhanced light-matter interaction strongly decreases the radiative carrier lifetimes \cite{Strauf2011}. As a consequence, the capture rates $S_{n/p}$ are expected to increase by some orders of magnitude such that the quantum system significantly couples back to its classical environment and contributes to current guiding. We suspect that in this case the predictions of the hybrid model differ clearly from a decoupled treatment.

An interesting extension of the system (\ref{eq: Poisson equation})--(\ref{eq: quantum master equation}) concerns reservoirs with different temperatures, as frequently studied in quantum thermodynamics \cite{Kosloff2013, Strasberg2017}. We are confident that it is possible to achieve a thermodynamically consistent coupling of the quantum master equation (\ref{eq: quantum master equation}) with energy transport models \cite{Albinus2002} or other transport models taking higher moments of the semi-classical Boltzmann equation \cite{Juengel2009} into account.
The latter extend the isothermal van Roosbroeck system by one or multiple heat flow equations that determine the spatial temperature distribution of the crystal lattice and the continuum carriers. The construction of the corresponding hybrid system should be analogous to the case considered in this paper. The essential difference is that the coupling of both subsystems involves spatially averaged thermodynamic forces instead of chemical potentials, e.g. $\langle \mu_c\rangle_w \to \langle \mu_c/T\rangle_w$ etc.
What might be interesting in the non-isothermal case is the impact of the quantum-classical interactions on the heat generation.

\section{Summary}

Nowadays, quantum optical technologies are on their way from the lab to real world applications.
To advance this development, device engineers will need simulation tools, which combine classical device physics with models from cavity quantum electrodynamics.
As a step on this route, we have presented a new modeling approach for the simulation
of single and few quantum dot devices.

By connecting semi-classical carrier transport
theory with a quantum master equation in Lindlad form, our approach
has lead to a hybrid quantum-classical system, that allows for a comprehensive
description of electrically driven quantum dot devices on multiple scales: It enables the
computation of the spatially resolved carrier transport
together with the calculation of quantum optical figures of merit (e.g.
photon generation rates, higher order correlation functions) in realistic
semiconductor structures in a unified way. This has been demonstrated by numerical
simulations of an electrical single-photon source based on
a single quantum dot. 
We have presented a thorough theoretical analysis
of the approach and showed that it guarantees the conservation of
charge and the consistency with the thermodynamic equilibrium. Finally,
we have proven that our hybrid quantum-classical system obeys the
second law of thermodynamics. 

We believe that our approach serves as a blueprint for the simulation of further quantum dot based photonic devices, in particular nanolasers.

\begin{acknowledgments}
The work of M.\,K. has been support by the Deutsche Forschungsgemeinschaft
(DFG) within the collaborative research center 787 \emph{Semiconductor
Nanophotonics} under grant B4. M.\,M. was supported by the ERC via
AdG 267802 \emph{AnaMultiScale}. The authors acknowledge valuable
discussions with H.-J.~Wünsche, U.~Bandelow, D.~Peschka and
A.~Mielke. The authors are grateful to the reviewer for the detailed and helpful comments.
\end{acknowledgments}

\appendix

\section{Boundary conditions\label{sec:Boundary-conditions}}

We assume a decomposition of the domain boundary
\[
\partial\Omega=\Big(\bigcup_{i}\Gamma_{i}\Big)\cup\partial\Omega_{N}
\]
into several ohmic contacts and artificial boundaries
of the device \cite{Selberherr1984}. On the
artificial boundaries $\partial\Omega_{N}$, we assume homogeneous Neumann conditions
\[
\mathbf{n}\cdot\nabla\psi=0,\quad\mathbf{n}\cdot\nabla\mu_{c}=0,\quad\mathbf{n}\cdot\nabla\mu_{v}=0,
\]
 where $\mathbf{n}$ denotes the outer normal
vector. The ohmic contacts are modeled by Dirichlet boundary conditions
\[
\psi=\psi_{\text{eq}}+U_{\text{appl},i},\quad\mu_{c}=\mu_{i},\quad\mu_{v}=\mu_{i},
\]
on $\Gamma_{i}$, where $U_{\text{appl},i}$ represents the applied
voltage at the $i$-th ohmic contact and $\mu_{i}=\mu_{\text{eq}}-qU_{\text{appl},i}$.
The value of the built-in potential $\psi_{\text{eq}}$ is obtained
from the local charge neutrality condition at the ohmic boundaries
and zeros bias conditions ($\mu_{i}\equiv\mu_{\text{eq}}$ $\forall i$)
\cite{Farrell2017}.

\section{Electrostatic field energy\label{sec:Electrostatic-field-energy}}

Following \cite{Albinus1996}, we split the electrostatic potential
\[
\psi=\psi_{\text{int}}+\psi_{\text{ext}}
\]
into an internal field $\psi_{\text{int}}$ generated by the internal
charge density and an external field $\psi_{\text{ext}}$, which arises
from the built-in doping profile and the applied voltages. Consequently,
the Poisson problem (\ref{eq: Poisson equation})  is
decomposed into
\begin{align*}
-\nabla\cdot\varepsilon\nabla\psi_{\text{int}} & =q\rho_{\text{int}},\\
-\nabla\cdot\varepsilon\nabla\psi_{\text{ext}} & =qC,
\end{align*}
such that the internal field $\psi_{\text{int}}=\psi_{\text{int}}\left(\rho_{\text{int}}\right)$
can be written as a functional of the total internal carrier density
\[
\rho_{\text{int}}=p-n+Q\left(\rho\right).
\]
On the domain boundaries it holds
\begin{align*}
\mathbf{n}\cdot\varepsilon\nabla\psi_{\text{int}} & =0 &  & \text{on }\partial\Omega_{N},\\
\psi_{\text{int}} & =0 &  & \text{on }\Gamma_{i},
\end{align*}
and
\begin{align*}
\mathbf{n}\cdot\varepsilon\nabla\psi_{\text{ext}} & =0 &  & \text{on }\partial\Omega_{N},\\
\psi_{\text{ext}} & =\psi_{\text{eq}}+U_{\text{appl},i} &  & \text{on }\Gamma_{i}.
\end{align*}
A variation of the internal carrier density $\rho_{\text{int}}\to\rho_{\text{int}}+a\delta\rho$
($0<a\ll1$ is a small parameter) in the interior of the domain yields
a variation of the electrostatic field $\delta\psi$ according to
\[
-\nabla\cdot\varepsilon\nabla\delta\psi=q\delta\rho\quad\text{on }\Omega
\]
with the same boundary conditions for $\delta\psi$ as for $\psi_{\text{int}}$
stated above. The variation of the internal energy given by Eq.~(\ref{eq: internal energy electric field})
leads to
\begin{align*}
U_{\psi}\left(\rho_{\text{int}}+a\delta\rho\right) & =U_{\psi}\left(\rho_{\text{int}}\right)\\
 & \phantom{=}+a\int_{\Omega}\mathrm{d}^{3}r\,\varepsilon\nabla\psi_{\text{int}}\left(\rho_{\text{int}}\right)\cdot\nabla\delta\psi\\
 & \phantom{=}+aq\int_{\Omega}\mathrm{d}^{3}r\,\delta\rho\psi_{\text{ext}}+\mathcal{O}\left(a^{2}\right).
\end{align*}
Finally, using the identity
\begin{align*}
\int_{\Omega}\mathrm{d}^{3}r\,\varepsilon\nabla\psi_{\text{int}}\left(\rho_{\text{int}}\right)\cdot\nabla\delta\psi & =q\int_{\Omega}\mathrm{d}^{3}r\,\psi_{\text{int}}\left(\rho_{\text{int}}\right)\delta\rho,
\end{align*}
one obtains the Gâteaux-derivative 
\begin{align*}
\lim_{a\to0} & \frac{U_{\psi}\left(\rho_{\text{int}}+a\delta\rho\right)-U_{\psi}\left(\rho_{\text{int}}\right)}{a}=\\
 & \quad\quad\quad=q\int_{\Omega}\mathrm{d}^{3}r\,\left(\psi_{\text{int}}\left(\rho_{\text{int}}\right)+\psi_{\text{ext}}\right)\delta\rho.
\end{align*}
The central feature of the field's internal energy expression Eq.~(\ref{eq: internal energy electric field})
is \cite{Albinus1996,Albinus2002}
\begin{equation}
\frac{\delta U_{\psi}}{\delta\rho}=q\text{\ensuremath{\psi}}.\label{eq: functional derivative of electric field energy}
\end{equation}

\section{Entropy production rate \label{sec:Entropy-production-rate}}

This section gives some details on the derivation of the expression
(\ref{eq: entropy production rate}) for the entropy production rate.
Starting from Eq.~(\ref{eq: entropy production (general)}), one
obtains by using Eq.~(\ref{eq: total thermodynamic potentials})
and (\ref{eq: classical entropy and energy density}) the entropy
production rate as
\begin{align*}
\frac{\mathrm{d}S_{\text{tot}}}{\mathrm{d}t} & =-\frac{1}{T}\int_{\Omega}\mathrm{d}^{3}r\,\left(\frac{\partial u_{\text{cl}}\left(n,p\right)}{\partial t}-T\frac{\partial s_{\text{cl}}\left(n,p\right)}{\partial t}\right)\\
 & \phantom{=}-\frac{1}{T}\mathrm{tr}\left(H\mathcal{L}\left(\rho;\chi_w\right)\right)-k_{B}\mathrm{tr}\left(\log\left(\rho\right)\mathcal{L}\left(\rho;\chi_w\right)\right)\\
 & \phantom{=}-\frac{1}{T}\frac{\mathrm{d}U_{\psi}}{\mathrm{d}t}+\sum_{i\geq1}\frac{\mu_{i}}{qT}\int_{\Gamma_{i}}\mathrm{d}\mathbf{A}\cdot\left(\mathbf{j}_{n}+\mathbf{j}_{p}\right).
\end{align*}
Taking the partial time derivatives, using the state equations (\ref{eq: carrier densities}),
Eq.~(\ref{eq: functional derivative of electric field energy}) and
\[
\frac{\mathrm{d}U_{\psi}}{\mathrm{d}t}=\int_{\Omega}\mathrm{d}^{3}r\,q\psi\frac{\partial\left(p-n+Q\left(\rho\right)\right)}{\partial t},
\]
we arrive at
\begin{align*}
\frac{\mathrm{d}S_{\text{tot}}}{\mathrm{d}t} & =-\frac{1}{T}\int_{\Omega}\mathrm{d}^{3}r\,\left(\mu_{c}\frac{\partial n}{\partial t}-\mu_{v}\frac{\partial p}{\partial t}\right)\\
 & \phantom{=}-\frac{1}{T}\mathrm{tr}\left(H\mathcal{L}\left(\rho;\chi_w\right)\right)-k_{B}\mathrm{tr}\left(\log\left(\rho\right)\mathcal{L}\left(\rho;\chi_w\right)\right)\\
 & \phantom{=}+\frac{q}{T}\left\langle \psi\right\rangle _{w}\mathrm{tr}\left(N\mathcal{L}\left(\rho;\chi_w\right)\right)\\
 & \phantom{=}+\sum_{i\geq1}\frac{\mu_{i}}{qT}\int_{\Gamma_{i}}\mathrm{d}\mathbf{A}\cdot\left(\mathbf{j}_{n}+\mathbf{j}_{p}\right),
\end{align*}
where we have explicitly used Eq.~(\ref{eq: charge density of the quantum system})
for the charge density of the quantum system. For different $Q\left(\rho\right)$
and multiple QDs, the calculation follows the same lines. With the
help of the carrier transport equations (\ref{eq: electron transport})--(\ref{eq: hole transport}),
the macroscopic capture rates (\ref{eq: loss terms}), partial integration
and the boundary conditions given in Appendix~\ref{sec:Boundary-conditions},
this is
\begin{align*}
\frac{\mathrm{d}S_{\text{tot}}}{\mathrm{d}t} & =\frac{1}{T}\int_{\Omega}\mathrm{d}^{3}r\,\left(\mu_{c}-\mu_{v}\right)R\\
 & \phantom{=}+\frac{1}{qT}\int_{\Omega}\mathrm{d}^{3}r\,\left(\mathbf{j}_{n}\cdot\nabla\mu_{c}+\mathbf{j}_{p}\cdot\nabla\mu_{v}\right)\\
 & \phantom{=}+\frac{1}{T}\left\langle \mu_{c}\right\rangle_{w}\mathrm{tr}\left(N \mathcal{D}_{e}(\rho;\chi_w)\right)\\
 & \phantom{=}+\frac{1}{T}\left\langle \mu_{v}\right\rangle_{w}\mathrm{tr}\left(N \mathcal{D}_{h}(\rho;\chi_w)\right)\\
 & \phantom{=}-\frac{1}{T}\mathrm{tr}\left(H\mathcal{D}\left(\rho;\chi_w\right)\right) -k_B\mathrm{tr}\left(\log\left(\rho\right)\mathcal{D}\left(\rho;\chi_w\right)\right)\\
 & \phantom{=}+\frac{q}{T}\left\langle \psi\right\rangle _{w}\mathrm{tr}\left(N\mathcal{D}\left(\rho;\chi_w\right)\right).
\end{align*}
In the above expression, the surface integrals have canceled out.
Using Eqns.~(\ref{eq: dissipator decomposition})--(\ref{eq: charge conservation of D0}),
one arrives at Eq.~(\ref{eq: entropy production rate}).

\section{Second law of thermodynamics \label{subsec:Positivity-of-the} }

In this section we proof the non-negativity of the entropy production
rate (\ref{eq: entropy production rate}) of the hybrid system (\ref{eq: Poisson equation})--(\ref{eq: quantum master equation}).
First, we introduce the (auxiliary) density matrices\begin{subequations}\label{eq: auxiliary stationary states}
\begin{align}
\rho_{0}^{\ast} & =\frac{1}{Z_{0}^{\ast}}e^{-\beta H},\label{eq: auxiliary stationary states 0}\\
\rho_{e}^{\ast}(\chi_w) & =\frac{1}{Z_{e}^{\ast}}e^{-\beta\left(H-\mu_{c}^{\text{eff}}(\chi_w)N\right)},\label{eq: auxiliary stationary states e}\\
\rho_{h}^{\ast}(\chi_w) & =\frac{1}{Z_{h}^{\ast}}e^{-\beta\left(H-\mu_{v}^{\text{eff}}(\chi_w)N\right)}\label{eq: auxiliary stationary states h}
\end{align}
\end{subequations}with $\mu_{c/v}^{\text{eff}}(\chi_w)=\left\langle \mu_{c/v}\right\rangle _{w}+q\left\langle \psi\right\rangle _{w}$.
Using Eq.~(\ref{eq: quantum detailed balance rates}) it can be shown
by direct calculation that $\mathcal{D}_{\nu}\left(\rho_{\nu}^{\ast}(\chi_w);\chi_w\right)=0$,
$\nu\in\left\{ 0,e,h\right\} $, for the dissipators given in Eq.~(\ref{eq: dissipator decomposition}).
Then, \textit{Spohn's inequaility} \cite{Spohn1978} states that
\begin{equation}
\mathrm{tr}\left(\left(\log\rho_{\nu}^{\ast}(\chi_w)-\log\rho\right)\mathcal{D}_{\nu}\left(\rho;\chi_w\right)\right)\geq0\label{eq: entropy production rate inequality}
\end{equation}
for $\nu\in\left\{ 0,e,h\right\} $. 

The entropy production rate Eq.~(\ref{eq: entropy production rate})
can be written in the form
\begin{align}
\frac{\mathrm{d}S_{\text{tot}}}{\mathrm{d}t} & =k_{B}\int_{\Omega}\mathrm{d}^{3}r\,\beta\left(\mu_{c}-\mu_{v}\right)\left(1-e^{-\beta\left(\mu_{c}-\mu_{v}\right)}\right)\sum_{j}r_{j}\nonumber \\
 & \phantom{=}+\frac{1}{q^{2}T}\int_{\Omega}\mathrm{d}^{3}r\,\left(\sigma_{n}\left|\nabla\mu_{c}\right|^{2}+\sigma_{p}\left|\nabla\mu_{v}\right|^{2}\right)\nonumber \\
 & \phantom{=}+k_{B}\mathrm{tr}\left(\left(\log{\rho_{0}^{\ast}}-\log{\rho}\right)\mathcal{D}_{0}\left(\rho;\chi_w\right)\right)\label{eq: entropy production rate (rearranged)}\\
 & \phantom{=}+k_{B}\mathrm{tr}\left(\left(\log{\rho_{e}^{\ast}(\chi_w)}-\log{\rho}\right)\mathcal{D}_{e}\left(\rho;\chi_w\right)\right)\nonumber \\
 & \phantom{=}+k_{B}\mathrm{tr}\left(\left(\log{\rho_{h}^{\ast}(\chi_w)}-\log{\rho}\right)\mathcal{D}_{h}\left(\rho;\chi_w\right)\right),\nonumber 
\end{align}
where we have used Eqns.~(\ref{eq: current densities}), (\ref{eq: auxiliary stationary states}),
the trace conservation property of the dissipator and a recombination
rate of the form (\ref{eq: recombination rate}). Here, $j$ labels the recombination channels and the functions $r_{j}=r_{j}\left(n,p,\psi\right)$
are non-negative by construction (cf. Appendix \ref{sec: Parameters and auxiliary - van Roosbroeck}). Using the inequalities (\ref{eq: entropy production rate inequality})
and $x\left(1-e^{-x}\right)\geq0$ $\forall x\in\mathbb{R}$, it is
easy to see that each line of Eq.~(\ref{eq: entropy production rate (rearranged)})
is non-negative. 

\section{Projection on eigenstates\label{sec:Projection-on-eigenstates}}

In order to obtain a system of ODEs from Eq.~(\ref{eq: quantum master equation}), it must be projected on a basis of the quantum system's Hilbert space. We use the eigenbasis of
the Hamiltonian $H$, for which we assume the spectral representation
\[
H=\sum_{k}\varepsilon_{k}\vert\varphi_{k}\rangle\langle\varphi_{k}\vert.
\]
For the sake of simplicity, we consider the energy spectrum $\left\{ \varepsilon_{k}\right\} $
to be non-degenerate here. Moreover, the Lamb-Shift contribution to $H$ is neglected. Then, the jump operators are projectors
between energy eigenstates $A_{\alpha}\to A_{i,j}=\vert\varphi_{i}\rangle\langle\varphi_{j}\vert$. The equations of motion for the diagonal elements of the density
matrix are obtained as
\begin{align*}
\partial_{t}\langle\varphi_{k}\vert\rho\vert\varphi_{k}\rangle & =\sum_{j}\left(\mathcal{M}_{k,j}\langle\varphi_{j}\vert\rho\vert\varphi_{j}\rangle-\mathcal{M}_{j,k}\langle\varphi_{k}\vert\rho\vert\varphi_{k}\rangle\right),
\end{align*}
whereas the off-diagonal elements $k\neq l$ obey
\begin{align*}
\partial_{t}\langle\varphi_{k}\vert\rho\vert\varphi_{l}\rangle= & -\frac{i}{\hbar}\left(\varepsilon_{k}-\varepsilon_{l}\right)\langle\varphi_{k}\vert\rho\vert\varphi_{l}\rangle\\
 & -\frac{1}{2}\sum_{j}\left(\mathcal{M}_{j,k}+\mathcal{M}_{j,l}\right)\langle\varphi_{k}\vert\rho\vert\varphi_{l}\rangle
\end{align*}
with the (non-negative) transition rate matrix elements
\begin{align*}
\mathcal{M}_{i,j} & =\gamma_{i,j}+\hat{\gamma}_{j,i}\\
 & =\gamma_{i,j}\left(1+e^{-\beta\left(\varepsilon_{j}-\varepsilon_{i}-\left(\left\langle \mu_{i,j}\right\rangle _{w}+q\left\langle \psi\right\rangle _{w}\right)\ell_{i,j}\right)}\right)\geq0.
\end{align*}
Obviously, in the case of non-degenerate energy spectra the diagonal
elements decouple from the off-diagonal elements. The off-diagonal
elements are fully decoupled each and show damped oscillations (dephasing).
This has important implications on the complexity of the
numerical simulations: Starting from the thermodynamic equilibrium state,
where only diagonal elements of the density matrix are occupied, the
dynamics never excite any off-diagonal elements (in the energy eigenbasis representation). Hence, the off-diagonal elements
can be omitted from the simulation. Thereby the
number of degrees of freedom of the quantum system grows only with
$N$ instead of $N^{2}$, where $N$ is the dimension of the (possibly
truncated) Hilbert space. However, if the spectrum of $H$ is degenerate, this feature is lost in general and one has to account for degenerate eigenstate coherences (i.e. off-diagonal elements contribute to the dynamics) \cite{Cuetara2016}.

\section{Parameters and auxiliary models\label{sec:Parameters-and-auxiliary}}

This section lists auxiliary models and parameters used in the numerical simulations
presented in Sec.~\ref{sec:Application}.

\subsection{Van Roosbroeck system} \label{sec: Parameters and auxiliary - van Roosbroeck}

We use GaAs parameters at $T=50\,\text{K}$. The effective masses
are $m_{e}^{\ast}=0.068\,m_{0}$, $m_{h}^{\ast}=0.503\,m_{0}$,
where $m_{0}$ denotes the (free) electron mass and the band edge
energies are taken as $E_{v}=0\,\text{eV}$ and $E_{c}=1.516\,\text{eV}$.
The (static) relative permittivity is set to $\varepsilon_{r}=12.9$,
the LO-phonon energy is $\hbar\omega_{\text{LO}}=36.5\,\text{meV}$
and the refractive index is $n_{r}=3.55$
(around $\text{950\,nm}$). The (net-)recombination rate in Eqns.~(\ref{eq: electron transport}),
(\ref{eq: hole transport}) is modeled as \cite{Selberherr1984,Farrell2017}
\begin{align}\label{eq: recombination rate}
\begin{aligned} 
R&=R_{\text{SRH}}+R_{\text{sp}}+R_{\text{Au}} \\
&= \left(1-e^{-\beta\left(\mu_{c}-\mu_{v}\right)}\right) \sum_j r_j(n,p,\psi)
\end{aligned}
\end{align}
where $j\in\{ \text{SRH},\text{sp},\text{Au} \}$ labels the different channels and
\begin{align*}
r_{\text{SRH}} & =\frac{np}{\tau_{p}\left(n+n_{d}\right)+\tau_{n}\left(p+p_{d}\right)},\\
r_{\text{sp}} & =Bnp,\\
r_{\text{Au}} & =\left(C_{\text{Au}}^{n}n+C_{\text{Au}}^{p}p\right)np
\end{align*}
with $n_{d}=ne^{\beta\left(E_{T}-q\psi-\mu_{c}\right)}$, $p_{d}=pe^{-\beta\left(E_{T}-q\psi-\mu_{v}\right)}$.
The recombination rates $r_j(n,p,\psi)$ of the individual channels are non-negative by construction.
The non-radiative life times are sensitive to the impurity concentration
and modeled via $\tau_{n/p}=\tau_{n/p,0}/\big(1+\big(\vert C\vert/C_{\text{ref}}\big)^{\gamma_{\text{SRH}}}\big)$
with $\tau_{n,0}=\tau_{p,0}=10\,\text{ns}$, $\gamma_{\text{SRH}}=1.72$
and $C_{\text{ref}}=9\times10^{17}\,\text{cm}^{-3}$ \cite{Palankovski2004}.
The trap energy level $E_{T}$ is assumed to be in the center of the
energy gap. The radiative recombination coefficient is taken as $B=1.06\times10^{-8}\,\text{cm}^{-3}\,\text{s}^{-1}$
and the Auger recombination coefficients are set to $C_{\text{Au}}^{n}=6\times10^{-30}\,\text{cm}^{-6}\,\text{s}^{-1}$,
$C_{\text{Au}}^{p}=1.6\times10^{-29}\,\text{cm}^{-6}\,\text{s}^{-1}$
\cite{Palankovski2004}. The carrier mobilities $M_{n/p}$ are taken
from the model given in Ref. \cite{Mnatsakanov2004}, which is reported to hold down to $T=50\,\text{K}$.
Despite the low temperatures, we assume complete ionization
due to the metal-insulator transition at heavy doping \cite{Mott1968}.

\subsection{Open quantum system}

The eigenenergies of the Hamiltonian (\ref{eq: application Hamiltonian})
are obtained from the parabolic/step-like confinement potential (relative
to the respective continuum band edge) $U_{\lambda}\left(r,z\right)=-U_{0}^{\lambda}\Theta\left(h/2-\left|z\right|\right)+\frac{1}{2}m_{\lambda}^{\ast}\omega_{\lambda,0}^{2}r^{2}$, $\lambda\in\left\{ e,h\right\} $, by solving the stationary Schr\"{o}dinger
equation at flat band conditions \cite{Wojs1996,Nielsen2004}. The
parameters for the InGaAs-QD are taken as $U_{e}=\text{350\,\text{meV}}$,
$U_{h}=\text{170\,\text{meV}}$, $m_{e}^{\ast}=0.067\,m_{0}$, $m_{h}^{\ast}=0.15\,m_{0}$
\cite{Nielsen2004} and $\hbar\omega_{e,0}=45.5\,\text{meV}$, $\hbar\omega_{h,0}=20.3\,\text{meV}$.
The QD height is assumed as $h=3\,\text{nm}$. For the computation
of the Coulomb matrix elements we set the background dielectric permittivity
to $\varepsilon_{r}=12.5$ \cite{Nielsen2004}. 

With the parameters above, the QD conduction band ground state $\varepsilon_{c}$
is found at $137.7\,\text{meV}$ below the continuum band edge and
the QD valence band ground state $\varepsilon_{v}$ is $44.7\,\text{meV}$
above the valence band edge. The Coulomb matrix elements are obtained
as $V_{c,c}=23.2\,\text{meV}$, $V_{v,v}=24.5\,\text{meV}$ and $V_{c,v}=23.7\,\text{meV}$.
The interband dipole moment is assumed as $d_{c,v}=q\times0.6\,\text{nm}$
and the Purcell factor is set to $P_{i,f}=1.8$ for all allowed optical
transitions. The emission energies are obtained around $1.31\,\text{eV}$
with radiative life times of approximately $1\,\text{ns}$ according
to Eq.~(\ref{eq: radiative decay rate}). The fitting parameters
in the carrier scattering rates are set to $\tau_{\text{LO}}^{e}=\tau_{\text{LO}}^{h}=10\,\text{ps}$,
$a_{\text{LO}}^{e}=25\,\text{meV}$, $a_{\text{LO}}^{h}=7\,\text{meV}$,
$\tau_{\text{Au}}^{\lambda,\lambda^{\prime}}=1\,\text{ps}$, $\gamma_{\text{Au}}^{\lambda,\lambda^{\prime}}=0.7$
(for all $\lambda,\lambda^{\prime}\in\left\{ e,h\right\} $), $n_{\text{Au}}^{\text{crit}}=1\times10^{19}\,\text{cm}^{-3}$
and $p_{\text{Au}}^{\text{crit}}=5\times10^{18}\,\text{cm}^{-3}$.

%

\end{document}